# Reconstruction of recurrent synaptic connectivity of thousands of neurons from simulated spiking activity

Yury V. Zaytsev[1,2,6] · Abigail Morrison[1,3,4] · Moritz Deger[5]

**Abstract** Dynamics and function of neuronal networks are determined by their synaptic connectivity. Current experimental methods to analyze synaptic network structure on the cellular level, however, cover only small fractions of functional neuronal circuits, typically without a simultaneous record of neuronal spiking activity. Here we present a method for the reconstruction of large recurrent neuronal networks from thousands of parallel spike train recordings. We employ maximum likelihood estimation of a generalized linear model of the spiking activity in continuous time. For this model the point process likelihood is concave, such that a global optimum of the parameters can be obtained by gradient ascent. Previous methods, including those of the same class, did not allow recurrent networks of that order of magnitude to be reconstructed due to prohibitive computational cost and numerical instabilities. We describe a minimal model that is optimized for large networks and an efficient scheme for its parallelized numerical optimization on generic computing clusters. For a simulated balanced random network of 1000 neurons, synaptic connectivity is recovered with a misclassification error rate of less than 1 % under ideal conditions. We show that the error rate remains low in a series of example cases under progressively less ideal conditions. Finally, we successfully reconstruct the connectivity of a hidden synfire chain that is embedded in a random network, which requires clustering of the network connectivity to reveal the synfire groups. Our results demonstrate how synaptic connectivity could potentially be inferred from large-scale parallel spike train recordings.

✉ Yury V. Zaytsev
yury@zaytsev.net

Abigail Morrison
morrison@fz-juelich.de

Moritz Deger
moritz.deger@epfl.ch

[1] Simulation Laboratory Neuroscience – Bernstein Facility for Simulation and Database Technology, Institute for Advanced Simulation, Jülich Aachen Research Alliance, Jülich Research Center, Jülich, Germany

[2] Faculty of Biology, Albert-Ludwig University of Freiburg, Freiburg im Breisgau, Germany

[3] Institute for Advanced Simulation (IAS-6), Theoretical Neuroscience & Institute of Neuroscience and Medicine (INM-6), Computational and Systems Neuroscience, Jülich Research Center and JARA, Jülich, Germany

[4] Institute of Cognitive Neuroscience, Faculty of Psychology, Ruhr-University Bochum, Bochum, Germany

[5] School of Life Sciences, Brain Mind Institute and School of Computer and Communication Sciences, École polytechnique fédérale de Lausanne, 1015 Lausanne EPFL, Switzerland

[6] Forschungszentrum Jülich GmbH, Jülich Supercomputing Center (JSC), 52425 Jülich, Germany

# 1 Introduction

The synaptic organization of neuronal networks is key to understanding the dynamics of brain circuits, and, eventually, to link them to higher level cognitive functions. A large body of work aims to address this challenge by developing experimental techniques which enable the reconstruction of the connections between neurons on the basis of anatomical or physiological evidence. Anatomically, synaptic connections may be identified using optical imaging or electron microscopy (Briggman et al. 2011; Bock et al. 2011), while physiological approaches rely on simultaneous recordings of individual neurons and the mutual influence of the spikes of one neuron on the membrane potential of the other (Perin et al. 2011; Boucsein et al. 2011). Substantial progress has been made in recent decades to increase the size of networks accessible by experimental methods, including the new promising macroscale and mesoscale connectivity mapping techniques (Chung et al. 2013; Oh et al. 2014). However, on the microscale of individual neurons, the practical limitations of these techniques mean that reliable reconstruction is currently only possible for neural circuits of up to dozens of cells.

Alternatively, the connectivity of neuronal networks can be inferred from parallel recordings of their spiking activity. Potentially, this enables the recovery of the connections in circuits of hundreds and thousands of cells. Recent technical achievements in conducting large-scale parallel recordings of neuronal dynamics, such as multi-electrode array technology for *in vivo* implantation (Hatsopoulos and Donoghue 2009; Ghane-Motlagh and Sawan 2013), micro-electrode dishes for recording the *in vitro* activity of acute brain slices and dissociated cell cultures (Nam and Wheeler 2011; Spira and Hai 2013), and optical imaging techniques (Grewe and Helmchen 2009; Lütcke et al. 2013; Ahrens et al. 2013), make this path even more compelling.

The main difficulty in the analysis of parallel recordings, though, lies in the interpretation of the results (Gerstein and Perkel 1969; Aertsen et al. 1989). On one hand, simple reduced models of network interactions are often unable to resolve ambiguous scenarios: a classic example of such ambiguity is a group of neurons that receives common input versus a mutually connected group of cells, which cannot be distinguished using pairwise cross-correlation analysis (Stevenson et al. 2008). On the other hand, obtaining reliable fits of complex large-scale models to the data presents both a methodological and computational challenge in itself (Chen et al. 2011; Song et al. 2013). At the same time, there are often considerable difficulties in directly relating the reconstructed connectivity matrices to measurable experimental quantities or model parameters. The resulting sets of connections are then regarded as "functional" or "effective" connectivity, terms lacking strict and universally accepted definitions, and not necessarily matching real anatomical connectivity, but still hoped to provide useful insights with respect to the interaction of the network elements (Horwitz 2003).

The desire to strike the balance between explanatory power, and analytical as well as numerical tractability, has fueled an ever growing interest in methods that go beyond simple linear regression analysis, but still remain highly efficient. Previous works show that *generalized linear models* (GLM) (McCullagh and Nelder 1989) of network spiking activity can indeed be efficiently estimated from experimental data (Truccolo et al. 2005; Okatan et al. 2005; Pillow et al. 2008; Stevenson et al. 2009; Gerwinn et al. 2010) (dealing with recordings of up to 20, 33, 27, 75 + 108 and 7 neurons respectively), and make it possible to recover the actual synaptic connectivity of small neuronal circuits ($N = 3$) (Gerhard et al. 2013). Scaling these approaches directly up to substantially larger networks of thousands of units, however, seemed not to be feasible due to the vast computational resources such a reconstruction would require.

In this work, we present a method to reconstruct the parameters of large-scale recurrent neuronal network models of $N \geq 1000$ elements, based on parameter estimation of a stochastic point process GLM using only observations of the spiking activity of the neurons. Provided with the knowledge of the probability $p(X|\theta)$ of a specific stochastic model yielding the observations $X$ given the parameters $\theta$, we maximize the likelihood function $L(\theta) = p(X|\theta)$ in order to identify a set of parameters $\theta$ resulting in an optimal agreement of the selected model with the observations $X$. This is a widespread technique known as *maximum likelihood estimation* (MLE) (Paninski 2004). If the underlying model is sufficiently detailed and is indeed appropriate to describe the observations, then not only can the parameters $\theta$ be related to the actual measurable features of the neuronal network that generated the data, but they also define a dynamic model of the neuronal network activity (also called a generative model). Such a model can be used to derive testable predictions, or conduct virtual experiments (simulations), which might otherwise have been impossible or impractical.

Due to the large number of parameters necessary to describe a network of $N \geq 1000$ neurons, the optimization of the likelihood $L(\theta)$ can only be performed efficiently for some of the possible GLMs of neuronal networks. In Section 2, we describe our optimized model, including a particular choice of nonlinearity and interaction kernels, which enables us to obtain closed forms and recurrence formulae which go beyond more general techniques previously reported in the literature. We additionally supply details about the numerical methods employed. In Section 3, we demonstrate the proposed technique on simulations of random balanced neuronal networks, and present reconstructions of the connectivity matrix consisting of $10^6$ possible synapses in sparsely connected recurrent networks of $N = 1000$ spiking neurons. Finally, we apply our method to a structured network. We recover a synfire chain embedded in a balanced network from recordings of spiking activity, in which no activations of the synfire chain were present, and demonstrate that the inferred model of this network supports the transmission of synfire activity when stimulated.

In the present study we focus on reconstructions of networks for which all spiking activity can be recorded. Whereas in experimental settings undersampling is to be expected – and we performed a basic assessment of how it would affect our reconstructions, see Appendix C – a thorough investigation of the consequences of undersampling for the classification performance of our techniques is out of scope. Similarly, when presenting these techniques we are initially concerned with activity which we can assume to be a sample of a multi-dimensional point process with constant parameters (i.e. neuronal excitability and synaptic interactions). In Section 4 we examine these limitations and propose how they could be relaxed in future studies.

## 2 Methods

This section provides detailed information on the method of network reconstruction we employ, including original amendments and adaptations. In Section 2.1 we introduce the likelihood of our network model to reproduce a given dataset of neuronal spike trains. This likelihood is the quantity which is subject to optimization. The specific formulation of the likelihood relies on a model of the spiking activity of the neurons, which is introduced in Section 2.2. To evaluate the likelihood and its gradient under that model efficiently, recursive formulae and closed form expressions are derived in Section 2.3. The subsequent sections describe how we handle synaptic transmission delays (Section 2.4) and how, in some cases, we employ regularization of the optimization problem (Section 2.5). Finally, Section 2.6 gives further details regarding the practical aspects of our highly parallelized implementation of the method.

### 2.1 Point process likelihood of generalized linear models

A statistical model that describes the activity of a network of $N$ neurons can be defined as an expression for the conditional probability $p(S|\vec{x})$ of observing an $N$-dimensional spike train (spike raster) $S$ for a given input signal $\vec{x}$, which may include external stimulation and/or previous activity of the network itself. Given all the inputs of a neuron, we assume that its probability of spiking is independent of the other neurons (conditional independence). This allows us to factorize $p(S|\vec{x}) = \prod_{i=1}^{N} p_i(S_i|\vec{x})$, where $p_i(S_i|\vec{x})$ is the probability that the $i$-th neuron, within the recording time $[T_0, T_1]$, produces a spike train $S_i$ conditioned on the input $\vec{x}$. Therefore, in what follows we focus on the probability $p_i(S_i|\vec{x})$ of a single neuron.

The activity of the individual nerve cells can be characterized by a stochastic GLM that postulates that two consecutive operations are performed by the neuron on its input. First, the dimensionality of the observable signal $\vec{x}$ is reduced by means of a linear transformation $\mathbf{K}_i$. This transformation models synaptic and dendritic filtering, input summation and leaky integration in the soma. The result $\mathbf{K}_i\vec{x}$ is a one-dimensional quantity that is analogous to the membrane potential of a point neuron model. Second, this transformed one-dimensional signal is fed into a nonlinear probabilistic spiking mechanism, which works by sampling from an inhomogeneous Poisson process with an instantaneous rate (conditional intensity function) given by $\lambda_i(t|\vec{x}) = f_i(\mathbf{K}_i\vec{x})$. Here, $f_i(\cdot)$ is a function that captures the nonlinear properties of the neuron. Both the linear filter $\mathbf{K}_i$ and the nonlinearity $f_i$ are specified by $\theta_i$, a set of parameters that describes the characteristics of the $i$-th neuron. The schematic of this model is shown in Fig. 1.

Based on these definitions, we may now introduce the natural logarithm $\mathcal{L}$ of the likelihood $L(\theta|S)$ and expand it as

$$\mathcal{L} = \log L(\theta|S) = \log \left[ \prod_{i=1}^{N} p_i(S_i|\vec{x}) \right]$$
$$= \sum_{i=1}^{N} \log p_i(S_i|\vec{x}) = \sum_{i=1}^{N} \mathcal{L}_i , \qquad (1)$$

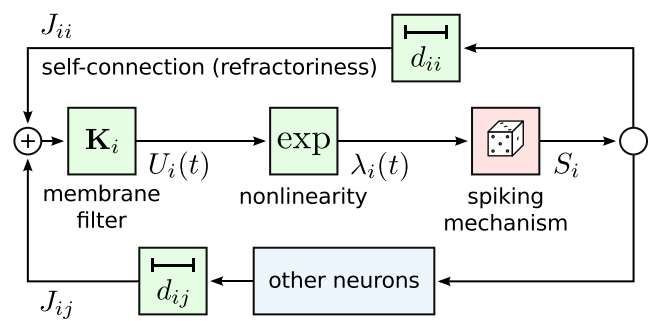

**Fig. 1** Schematic of the point process generalized linear model (PP GLM) of a recurrent spiking neuronal network. In this model, the spike trains $\vec{x}$ from the neurons in the network, after incurring transmission delays $d_{ij}$, pass through a linear filtering stage $\mathbf{K}_i$. The resulting (pseudo) membrane potential $U_i(t)$ is fed into a nonlinear link function $f_i(\cdot) = \exp(\cdot)$, which transforms it into the conditional intensity function $\lambda_i(t)$. The latter drives the probabilistic spiking mechanism that generates an output spike train $S_i$ for the $i$-th neuron. Note that this spike train is then also fed back as an input to the neuron itself via a "self-connection" in order to model its refractory, post-spike properties

where the observation (previously called $X$) is the spike raster $S$. In the last step of Eq. (1) we have introduced the single neuron log-likelihood $\mathcal{L}_i = \log p_i(S_i|\vec{x})$.

Let us now compute the probability that an inhomogeneous Poisson process with intensity $\lambda_i(t)$ produces the spike train $S_i = \{t_{i,k}\}$, $1 \leq k \leq q_i$, where $T_0 \leq t_{i,k} \leq T_1$ and $q_i$ is the number of spikes of the $i$-th neuron. This probability is (Brillinger 1988)

$$p_i(S_i|\vec{x}) = e^{-\int_{t_{i,q_i}}^{T_1} \lambda_i(t)dt} \prod_{k=1}^{q_i} e^{-\int_{t_{i,k-1}}^{t_{i,k}} \lambda_i(t)dt} \lambda(t_{i,k})$$

$$= e^{-\int_{T_0}^{T_1} \lambda_i(t)dt} \prod_{k=1}^{q_i} \lambda(t_{i,k})$$

with $t_{i,0} = T_0$. Here, for each spike time $t_{i,k}$, we multiply the (survival) probabilities $e^{-\int_{t_{i,k-1}}^{t_{i,k}} \lambda_i(t)dt}$ of not producing a spike in $(t_{i,k-1}, t_{i,k})$ with the intensity $\lambda_i(t_{i,k})$ at $t_{i,k}$. Finally, we factor in the probability $e^{-\int_{t_{i,q_i}}^{T_1} \lambda_i(t)dt}$ of not producing a spike in the recording time $(t_{i,q_i}, T_1]$, which remains after the last spike. The function $L_i(\theta_i|S_i) = p_i(S_i|\vec{x})$ is known as the point process likelihood (Snyder and Miller 1991).

Taking the logarithm yields the log-likelihood function

$$\mathcal{L}_i(\theta_i) = \sum_{k=1}^{q_i} \log \lambda_i(t_{i,k}) - \int_{T_0}^{T_1} \lambda_i(t)dt, \tag{2}$$

where the sum runs over all spikes $1 \leq k \leq q_i$ of the $i$-th neuron. The first term of this expression rewards high intensity at times $t_{i,k}$ when the spikes of the $i$-th neuron have been emitted, and the second term penalizes high intensity when no spikes have been observed. Different numbers of spikes $q_i$ render the absolute values of $\mathcal{L}_i$ difficult to compare among different neurons, but play no role when maximizing $\mathcal{L}_i$ with respect to $\theta_i$.

## 2.2 Conditional intensity model for a recurrent neural network

In order to investigate the recurrent aspects of the dynamics of the system, we define the observable input signal $\vec{x}$ for each neuron as the history of spikes recorded in the network up to a given point in time, including the spikes of the $i$-th neuron itself (which are used to model the refractory properties of the neuron). It is possible to include external inputs in this formulation, however this is not an option that we have pursued in the current work. Below follows a detailed discussion of the different components of the model as presented in Fig. 1.

For simplicity, we assume that the effect of each incoming spike can be modeled as an instantaneous current injection. The spike train $S_j$ of the $j$-th neuron as a function of time is expressed as $s_j(t) = \sum_{k=1}^{q_j} \delta(t - t_{j,k})$, where $t_{j,k}$ is the $k$-th spike of the $j$-th neuron. Each spike then elicits an exponential post-synaptic response in the neuron, due to the filtering properties of the membrane, $h_i(t) = H(t) \exp\{-t/\tau_i\}$, where $t$ is the time since spike arrival, $\tau_i$ is the membrane time constant of the neuron, and $H(x) = \{1$ if $x \geq 0$, else $0\}$ is the Heaviside function, which ensures the causal relationship between the stimulation and the response. Note that while the propagation of spikes is assumed to happen instantaneously in the formulation above, the incorporation of delays will be discussed in detail later in Section 2.4.

We may now define the linear dimensionality-reducing transformation $U_i(t) = \mathbf{K}_i \vec{x}(t)$ as

$$U_i(t) = J_{i0} + \sum_{j=1}^{N} J_{ij}(h_i * s_j)(t), \tag{3}$$

where $*$ denotes the convolution operation,

$$(h_i * s_j)(t) = \int_{-\infty}^{\infty} h_i(t - u)s_j(u)du.$$

The baseline potential $J_{i0}$ will be used later to set a base level of activity of the unit in the absence of inputs. Differentiation of Eq. (3) yields the first-order ordinary differential equation of the leaky integrator

$$\frac{d}{dt}U_i(t) = -\frac{1}{\tau_i}(U_i(t) - J_{i0}) + \sum_{j=1}^{N} J_{ij}s_j(t). \tag{4}$$

Hence $U_i(t)$ can be interpreted as the membrane potential of the $i$-th neuron, while $J$ is the synaptic connectivity matrix and each of its elements $J_{ij}$ denotes the strength (synaptic weight) of the connection from the $j$-th to the $i$-th neuron. Due to its simplicity, Eq. (4) leads to highly efficient algorithms (discussed in Section 2.3 and 2.6) to evaluate the membrane potential and the conditional intensity function of the neurons, beyond previously reported more general parallelization techniques (Chu et al. 2006). The membrane potential and the intensity are, in turn, needed to compute the values of the likelihood function and its gradient.

In Eqs. (3) and (4), positive and negative values of $J_{ij}$ correspond to excitatory and inhibitory connections respectively, and zero values denote the lack of a connection between two cells. Note that as formulated, this model does not ensure compliance with Dale's law (according to which each neuron can form synapses of only one type). However, we will show that this is an essentially negligible source of errors in the reconstructions presented below.

Further, we choose a specific type of the nonlinearity $f(u) = \exp\{u\}$, such that

$$\lambda_i(t) = \exp\{U_i(t)/\delta u\}. \tag{5}$$

In this expression, the scalar $\delta u > 0$ can be considered as the inverse "gain" of the nonlinearity. In the derivations that follow we will assume $\delta u = 1$ in order to simplify the expressions without loss of the generality, as different gains can be accommodated by rescaling the synapse weights $J_{ij}$ and the baseline potential $J_{i0}$ accordingly. In the absence of input spikes, $U_i = J_{i0}$, which leads to the base rate

$$c_i = \exp\{J_{i0}/\delta u\}. \tag{6}$$

It is worth mentioning that the base rate can effectively constitute a "sink" for spiking activity that cannot be explained by the recurrent network dynamics, such as external stimulation that has not been included in the present model, or missing inputs from unobserved neurons due to incomplete observations of the network (undersampling).

The model as formulated above is similar to the widely used cascade LNP model (Simoncelli et al. 2004), but in addition to the activity of the other cells in the ensemble, it also incorporates the spiking history of the neuron itself through its self-connection $J_{ii}$. An intuitive biological interpretation of this class of models, also known as the spike-response model with escape noise, in relation to the conventional integrate-and-fire model is given in Brillinger (1988) and Gerstner et al. (2014). Here, in contrast to the approaches taken in previous studies (Song et al. 2013; Citi et al. 2014; Ramirez and Paninski 2014), we drastically simplify both the conditional intensity model for a single neuron and the interaction kernels. This makes the numerics in our method amenable to a highly efficient implementation as discussed in Section 2.6.

Given that $f_i(\cdot)$ is both a convex and log-concave function of $U_i = \mathbf{K}_i \vec{x}$, and the space of possible $\{\mathbf{K}_i\}$ is convex, it can be shown that the log-likelihood function of such problems is concave and does not have any non-global local extrema (Paninski 2004). Thus the log-likelihood function $\mathcal{L}_i$ of the model as formulated above is concave in $\theta_i \subset \{J_{ij}\}_{0 \leq j \leq N}$ (note, however, that $\tau_i$ is not included in $\theta_i$; the recovery of the time constants is addressed separately). A proof of the concavity of $\mathcal{L}_i$ for our specific choice of kernels and link function is given in Appendix A. Since the sum of concave functions is again concave, the full log-likelihood $\mathcal{L} = \sum_{i=1}^{N} \mathcal{L}_i$ is concave as well. Consequently, there exists a unique set of parameters $\theta$ that characterize the network model that is most likely to exhibit a given recorded activity. These parameters $\theta$ can be efficiently identified via gradient ascent based nonlinear optimization methods applied to $\mathcal{L}$. Moreover, due to the separability of $\mathcal{L}$ (1), in order to recover $\theta = \{\theta_i\}$, one can maximize the individual log-likelihood functions $\mathcal{L}_i$ for each recorded unit, instead of maximizing the complete log-likelihood function $\mathcal{L}$.

Since the experimental techniques to obtain simultaneous recordings of thousands of units are becoming increasingly accessible, in this work we are targeting $N \geq 1000$. However, even if the number of variables is reduced from the $\sim \mathcal{O}(N^2 = 10^6)$ required for the complete log-likelihood function to the $\sim \mathcal{O}(N = 10^3)$ required for the log-likelihood function of an individual neuron, this is still a high-dimensional convex optimization problem. It can only be solved in practice using gradient based methods, for which the analytical closed form expressions for the log-likelihood function and its gradient are both available, and amenable to efficient evaluation. In the following we derive these expressions for the postulated model.

### 2.3 Closed form expressions

Let us consider the log-likelihood $\mathcal{L}_i$ for an individual neuron; recall that the variable part of Eq. (2) consists of two terms:

$$\mathcal{L}_i = \underbrace{\sum_{k=1}^{q_i} \log \lambda_i(t_{i,k})}_{\mathcal{L}_i^{\Sigma}} - \underbrace{\int_{T_0}^{T_1} \lambda_i(t)dt}_{\mathcal{L}_i^{\int}}. \tag{7}$$

Observe that given a closed form for $U_i(t)$, computing $\mathcal{L}_i^{\Sigma}$ is a matter of a simple algebraic substitution, while the efficiency of computing $\mathcal{L}_i^{\int}$ depends on whether it is possible to find this primitive analytically.

#### 2.3.1 Recurrence formula for the membrane potential

By design, our particular choice of $\mathbf{K}_i$ (exponential postsynaptic potential plus baseline potential) allows us to obtain the required closed form for $U_i(t)$ because it obeys the leaky integrator dynamics (4). The solution of Eq. (4) from $t_k$ to $t$ in the absence of input spikes $s_j(t)$ is $U_i(t) = (U_i(t_k) - J_{i0})\exp\left\{-\frac{t-t_k}{\tau_i}\right\} + J_{i0}$. This expression is valid at any time $t$ between two consecutive observed spikes $t_k, t_{k+1} \in S$, where $S = \{t_k\}$ is the (ordered) set of all recorded spikes of the network. At the borders of each of those intervals, the value of $U_i(t_{k+1})$ is increased by the contribution of the corresponding incoming spike:

$$U_i(t_{k+1}) = (U_i(t_k) - J_{i0})e^{-\frac{t_{k+1}-t_k}{\tau_i}} + J_{i0} + J_{ij}, \tag{8}$$

where the index $j$ refers to the neuron that emitted a spike at time $t_{k+1}$; if spikes from multiple neurons $j^{1,2,3,...}$ arrive at time $t_{k+1}$, the contributions $J_{ij^{1,2,3,...}}$ have to be added. We will refer to Eq. (8) as the key recurrence formula in the following.

The formula (8) for $U_i(t_{k+1})$ makes it possible to find the value of the membrane potential of the neuron at the spike time $t_{k+1}$ given the previous value at time $t_k$ by computing only one exponential function. It is substantially more efficient in terms of computation than naively summing up the contributions from all spikes that happened at $t < t_k$ for each point in time $t_k$. In particular, for kernels with infinite memory like the exponential kernels $h_i(t)$ employed here, the recurrence formula (8) is crucial to avoid an explosion of the computational costs when evaluating the log-likelihood on large datasets in continuous time.

### 2.3.2 Evaluating the likelihood

Taking these considerations into account, the integral over the duration of the recording $\mathcal{L}_i^{\int}$ in Eq. (7) can be broken down into a sum of integrals from $t_k$ to $t_{k+1}$:

$$
\begin{aligned}
\mathcal{L}_i^{\int} &= \sum_{k=0}^{q+1} \int_{t_k}^{t_{k+1}} \lambda_i(t) dt \\
&= c_i \sum_{k=0}^{q+1} \int_{t_k}^{t_{k+1}} \exp\left\{ (U_i(t_k) - J_{i0}) e^{-\frac{t-t_k}{\tau_i}} \right\} dt ,
\end{aligned} \tag{9}
$$

where $q = \sum_{i=1}^{N} q_i$ is the total number of recorded spikes, $t_1, \ldots, t_q$ are the spike times, and $t_0 = T_0$ and $t_{q+1} = T_1$ are the start and end of the recording. The integral contained here has a known closed form, so

$$
\mathcal{L}_i^{\int} = -c_i \tau_i \sum_{k=0}^{q+1} \text{Ei}\left( (U_i(t_k) - J_{i0}) e^{-\frac{t-t_k}{\tau_i}} \right) \Big|_{t_k}^{t_{k+1}} , \tag{10}
$$

where $\text{Ei}(x)$ is a special function (exponential integral) defined as $\text{Ei}(x) = -\int_{-x}^{\infty} \frac{e^{-t}}{t} dt$ for real nonzero values of $x$. For a proof of the equivalence of Eqs. (9) and (10) see Appendix B; the numerical computation of this function is discussed below in Section 2.6.1. The summands of Eq. (10) are independent, and therefore the evaluation of $\mathcal{L}_i^{\int}$ lends itself to trivial parallelization.

### 2.3.3 Evaluating the gradient

It now remains to find an efficient way to compute the gradient of the log-likelihood function. The performance at this point is likewise important, or even more so for large $N$, since $\mathcal{L}_i$ has $\mathcal{O}(N)$ partial derivatives that all need to be evaluated at each step of the optimization. The parameters of $\mathcal{L}_i$ are $\theta_i = (J_{i0}, \ldots J_{iN})$. For convenience, let us first introduce the terms

$$
v_{ij}(t) = \frac{\partial}{\partial J_{ij}} U_i(t) = \begin{cases} j \geq 1 : (h_i * s_j)(t) \\ j = 0 : 1 \end{cases} , \tag{11}
$$

which, for $j \geq 1$, can be interpreted as the putative response of the $i$-th neuron to the input spikes from the $j$-th neuron, that is going to be scaled by $J_{ij}$, cf. (3). The derivatives of $\mathcal{L}_i$ (7) with respect to $J_{ij}$ can then be expressed as

$$
\frac{\partial}{\partial J_{ij}} \mathcal{L}_i = \underbrace{\sum_{k=1}^{q_i} v_{ij}(t_{i,k})}_{\partial_{ij}^{\Sigma}} - \underbrace{\int_{T_0}^{T_1} \lambda_i(t) v_{ij}(t) dt}_{\partial_{ij}^{\int}} . \tag{12}
$$

Here, $q_i$ is the number of spikes of the $i$-th neuron, and $\{t_{i,k}\} = S_i$ are the points in time when the $i$-th neuron emitted a spike. For $j = 0$, Eq. (12) becomes $\frac{\partial}{\partial J_{i0}} \mathcal{L}_i = q_i - \mathcal{L}_i^{\int}$. This means that at a maximum of $\mathcal{L}_i$, the baseline potential $J_{i0}$ (and so the base rate $c_i$ (6)) is set such that the number of spikes $q_i$ equals the expected total number of spikes of the GLM, $\mathcal{L}_i^{\int}$. Further, in order to evaluate (12) for the cases when $j \geq 1$, we have defined the symbols $\partial_{ij}^{\Sigma}$ and $\partial_{ij}^{\int}$ analogous to Eq. (7).

The values $v_{ij}(t_{i,k})$ for $j \geq 1$ can be obtained using a recurrence formula just like for the membrane potential $U_i(t_k)$ (8); in fact, $v_{ij}(t) = \frac{\partial}{\partial J_{ij}} U_i(t)$, cf. (11). Hence, $v_{ij}(t)$ obeys leaky integrator dynamics like $U_i(t)$, which can be obtained by differentiating Eq. (4) by $J_{ij}$ and reinserting Eq. (11). Accordingly, $v_{ij}(t)$ decays exponentially in between spikes $v_{ij}(t_{j,k} < t < t_{j,k+1}) = v_{ij}(t_k) \exp\left\{ -\frac{t-t_{j,k}}{\tau_i} \right\}$, and we find the recurrence formula

$$
v_{ij}(t_{j,k+1}) = v_{ij}(t_{j,k}) e^{-\frac{t_{j,k+1}-t_{j,k}}{\tau_i}} + 1 . \tag{13}
$$

Individual values within these intervals can be computed in parallel independently from each other. Summing up all $v_{ij}(t_{i,k})$ then yields $\partial_{ij}^{\Sigma}$.

It is also important to mention that $v_{ij}(t)$ (11) and, consequently, $\partial_{ij}^{\Sigma}$ in Eq. (12) do not depend on parameters $\theta_i$ and therefore need only be computed once at the beginning of the optimization. However, even though we can use the formula $U_i(t) = \sum_{j=0}^{N} J_{ij} v_{ij}(t)$, for large $N$ it is more expensive to compute $U_i(t)$ by summing up weighted contributions of $v_{ij}(t)$ than by using Eq. (8) as explained above.

Making use of the recurrence formulae for $U_i(t)$ (8) and $v_{ij}(t)$ (13), the closed form of $\partial_{ij}^{\int}$ in Eq. (12) can be expressed as follows:

$$
\begin{aligned}
\partial_{ij}^{\int} &= c_i \sum_{k=0}^{q+1} v_{ij}(t_k) \int_{t_k}^{t_{k+1}} e^{-\frac{t-t_k}{\tau_i}} e^{(U_i(t_k) - J_{i0}) \exp\left\{ -\frac{t-t_k}{\tau_i} \right\}} dt \\
&= -c_i \tau_i \sum_{k=0}^{q+1} \frac{v_{ij}(t_k)}{U_i(t_k) - J_{i0}} e^{(U_i(t_k) - J_{i0}) \exp\left\{ -\frac{t-t_k}{\tau_i} \right\}} \Big|_{t_k}^{t_{k+1}}
\end{aligned} \tag{14}
$$

where, as in Eq. (10), $q = \sum_{i=1}^{N} q_i$ is the total number of recorded spikes, $t_1, \ldots, t_q$ are the spike times, $t_0 = T_0$ and $t_{q+1} = T_1$ are the start and end of the recording. Unlike $\partial_{ij}^{\Sigma}$, this expression needs to be re-evaluated at every optimization step, but as with Eq. (10), the elements of the sum are independent from each other and can therefore also be efficiently parallelized.

## 2.4 Handling transmission delays

In the discussion above, the communication of spikes between the neurons was implicitly assumed to happen instantaneously. Of course, in reality spikes incur transmission delays, which strongly affects the dynamics of the network.

Fortunately, the effects of combined synaptic and axonal delays can be easily incorporated into the described model: thanks to the separability property, we can optimize the parameters for each neuron independently, and feed every optimization for different neurons with its own modified dataset, containing the incoming spike times from other neurons arriving as the target neuron actually received them.

Therefore, given an effective delay matrix $D$, it is only necessary to shift each spike train $S_j$ in the recorded raster $S$ by the corresponding delay at the beginning of the optimization for the $i$-th neuron, such that the membrane potential of this neuron is affected at the point in time when the incoming spikes from the $j$-th neuron have reached their target, and not immediately as they were fired (and recorded):

$$S_j = \{t_{j,k}\} \rightarrow \widehat{S_j} = \{\widehat{t_{j,k}} = t_{j,k} + D_{ij}\}. \tag{15}$$

The transformation above has to be applied with one exception: the elements of the sum in $\mathcal{L}_i^\Sigma$ (and, accordingly, $\partial_{ij}^\Sigma$) have to be evaluated at time points $S_i$ when the $i$-th neuron actually produced a spike, and not at time points $\widehat{S_i} = S_i + D_{ii}$, when this spike has reached the neuron through the "self-connection" and provoked a depression of its membrane potential, which models the refractory properties of the neuron.

In other words, in order to correctly evaluate the expressions Eqs. (7) and (12) while taking into account transmission delays, one must compute the values of $\widehat{U}_i(t)$ and $\widehat{v}_{ij}(t)$ using the *modified* raster $\widehat{S}$, but at time points $S_i$ of the *original* raster $S$, and substitute these values in the elements of the sums $\mathcal{L}_i^\Sigma$ and $\partial_{ij}^\Sigma$ respectively, instead of summing up the elements taken at times $\widehat{S}_i$. In the following, we omit the "hats" for notational convenience.

## 2.5 Regularization of the model

Substantial improvements in the quality of the network reconstruction can be achieved if the model presented above is subjected to standard regularization techniques. These techniques enhance the accuracy of the inference procedure by integrating additional prior knowledge about the system into the optimization process (Meinshausen and Bühlmann 2006; Ravikumar et al. 2010). For instance, we can impose box constraints on reasonable values of the synaptic connection matrix $J_{ij}$ or base rates $c_i$, and complement this with a choice of more sophisticated methods, such as $\ell_1$ or $\ell_2$ regularization, exploiting assumed sparsity or smoothness of the expected result, respectively (Chen et al. 2011).

In particular, $\ell_1$ regularization (Tibshirani 1996) has a straightforward Bayesian interpretation in our setting: by penalizing the log-likelihood function (2) with the sum of the absolute values of the synaptic weights $J_{ij}$, we impose a sparsity-inducing Laplace prior on the sought-for solution, thereby performing a *maximum a posteriori* (MAP) estimation. Here the strength of the penalty $\alpha$ reflects the firmness of our belief in the sparseness of the network connectivity:

$$\widetilde{\mathcal{L}}_i \sim \mathcal{L}_i^\Sigma - \mathcal{L}_i^f - \alpha \sum_{\substack{j=1 \\ j \neq i}}^N |J_{ij}|. \tag{16}$$

Possible overfitting due to an inadequate choice of the regularization parameter $\alpha$ can be prevented by separating the dataset into two parts to cross-validate the recovered synaptic weights, and, in the case that the available data is too scarce, more elaborate techniques such as $K$-fold cross-validation and other cross-validation types (Kohavi 1995) can be employed.

## 2.6 Practical implementation

The mathematical components described above make it possible to reproduce our estimation procedure. However, we found that without employing additional numerical methods, a naive implementation would be way too slow for practical use. In the following we outline the techniques that helped us to boost the optimization speed by many orders of magnitude, bringing the computational requirements to perform estimations of the connectivity for the networks of $N \sim \mathcal{O}(10^3)$ neurons into a practical range for plausible amounts of experimental data.

### 2.6.1 Efficient evaluation

From the computational perspective, a program that performs the parameter estimation would typically consist of a nonlinear optimization routine, which is provided with callback procedures that are repeatedly called in order to evaluate the objective function (2) and its gradient (12) for any given set of parameters. Hence, the cornerstone guiding principle to achieve best performance is to carefully consider the CPU time versus memory consumption trade-offs, and cache as many values for these callbacks as feasible.

As the values of $U_i(t)$ for $S = \{t_k\}$ (all spikes of the network) are needed in order to evaluate both the log-likelihood function and its gradient, it makes sense to pre-compute these values at the beginning of the optimization step. Additionally, as previously noted, the values of $v_{ij}(t)$ do not depend on the parameters $\theta$, and therefore both $v_{ij}(t)$ and $\partial_{ij}^{\Sigma}$ can be pre-computed during the first optimization step, and re-used in all subsequent steps. Likewise, it is important to consider the costs of calculating transcendental functions; whereas they might seem negligible at the first sight, the time taken to compute some $10^{10}$ exponentials every step is considerable. Therefore, pre-computing the values of all sub-expressions that do not depend on the parameters, and, in particular, $\xi_{t_k} = \exp\{-(t_{k+1}-t_k)/\tau_i\}$ is another possibility to save large amounts of CPU time.

In any case, we recommend using iterative profiling in order to select the relevant optimization targets to add each next level of caching, since, as a general rule, the more caches there are, the more complicated and error-prone it is to keep them consistent and up to date with respect to the changes in parameters. Additionally, this avoids the situations when a sizeable amount of work is invested only to gain minor improvements in speed, due to runtime actually being dominated by different code paths than anticipated.

We observed that the optimization algorithms are (unsurprisingly) sensitive to the precision of the evaluation of the objective function and its gradient, and especially to the consistency between the two. Therefore we rejected using numerical approximations to the gradient, such as values computed using the central differences formula, and employed analytically derived expressions instead. We have also found that better precision of the objective function leads to faster convergence. This particularly concerns the accurate approximation of the exponential integral in Eq. (10). In general, finding an efficient method to evaluate Ei($x$), which is a crucial part of Eq. (10), poses a significant computational challenge. However, high-quality rational approximations exist in the literature (Cody and Thacher 1969), which make it as fast as evaluating low-order polynomials. In our implementation, we rely on the approximations devised by John Maddock using a custom Remez code, which are part of the Boost C++ library.[1] These approximations are not only highly accurate, but also the fastest that are available to us.

### 2.6.2 Parallelization and distribution

As the sweeping growth of the clock speeds in the last couple of decades seems to have saturated, the focus is increasingly shifting towards increasing parallelism, and nowadays multicore CPUs are a *de facto* standard, rather than rare marvels. Therefore, suitability for parallelization is becoming a critical feature to discriminate the algorithms that are appropriate for large-scale data analysis. In this section we discuss the parallelization strategies applicable to the model described above.

Owing to the separability of the problem, the highest level approach to parallelize the execution of the optimization is to launch several estimations for different neurons in parallel. This results in a perfect scaling for $N_t \leq N$, where $N_t$ is the number of simultaneously executed hardware threads. This is clearly a very attractive option due to the relative simplicity of implementation, however, its practical applicability is limited by the amount of the available memory per thread, which quickly becomes a bottleneck for larger networks and bigger amounts of data.

A slightly lower-level method is to identify independent elements in the formulae that need to be evaluated at every step of the optimization, and divide this work among several threads within one running process. The summands of $\mathcal{L}_i^{\Sigma}$, $\mathcal{L}_i^f$, $\partial_{ij}^{\Sigma}$ and $\partial_{ij}^f$ as defined in Eqs. (7), (10), (12) and (14) are all amenable to that kind of processing. This approach is advantageous to utilize all usable threads from within one process, but its scalability is limited by both the amount of the available memory on a single compute node (as above), and the serial part of the computations, which cannot be parallelized. In our model, it is mainly the calculation of the membrane potential $U_i(t)$ (8) and the membrane responses $v_{ij}(t)$, because each value in the recurrence formulae depends on the previous one. The membrane responses $v_{ij}(t)$ are less of a problem, since they can be pre-computed at the beginning of the optimization as explained above, if one is willing to trade memory consumption for performance. Alternatively, $v_{ij}(t)$ can be computed in parallel, which can be faster than fetching the results from memory for a very high number of threads and low memory bandwidth.

We have also explored the possibility of distributing the estimation across several compute nodes, which is not only necessary in order to utilize larger numbers of threads than available on one node, but also allows the computation to make use of the additional memory when the problem gets too large to fit into one machine's RAM. The most straightforward distribution scheme is to designate one process (rank) to perform serial computations required for every optimization step, broadcast the results and parameters to other ranks, have them do their share of the computations, and, finally, collect the results. The biggest advantage of this scheme lies in its ease of implementation: the communication pattern is very clear, and the code can largely remain unchanged except for the need of a few additional functions to distribute and collect the data.

In our implementation, we performed the calculation of the membrane potential $U_i(t)$, the log-likelihood function $\mathcal{L}_i$ and $\partial \mathcal{L}_i / \partial J_{i0}$ on Rank 0, and evenly divided the work to compute $\partial \mathcal{L}_i / \partial J_{ij}$, $j \geq 1$, among all other ranks. This system scales (almost) linearly up to the point when the amount of

---

time needed to perform the computations on Rank 0 exceeds the amount of time it takes to compute the gradient distributed to all other ranks. Since it takes several orders of magnitude more time to calculate $\mathcal{L}_i$ than $\partial \mathcal{L}_i / \partial J_{ij}$, we have found that for $N = 1000$ we can easily distribute each single task up to $N_r = 10 \ldots 20$ ranks.

For production estimations, we combined all three approaches outlined above. The highest level of parallelization was left up to the batch system: for each estimation, we generated and submitted the job scripts for every neuron and let the scheduler optimally backfill the queue. The code was run with $N_t = 8 \ldots 16$, depending on the amount of hardware threads available per processor, and $N_r = 10 \ldots 20$, depending on the amount of available memory per processor and the requirements of the particular estimations. For estimations of size $N = 1000$, this hybrid approach allowed us to scale almost linearly up to $\mathcal{O}(N_t \times N_r \times N = 10^5)$ cores.

In this context, it becomes clear why not only the convexity, but also the separability property of the optimization problem discussed in Section 2.2 is crucial to our model. In a typical estimation, as described in Section 3, 1 hour recording of $N = 1000$ neurons spiking at $\sim 5 \, \mathrm{s}^{-1}$ would contain $\sim 10^7$ spikes, so the intermediary data to be held in RAM during the optimization would need around $\sim 10^{14} = 10 \times 10^7 \times (10^3)^2$ bytes or 100 TB of storage capacity. This calculation assumes that the main contribution comes from the pre-computed matrix of $v_{ij}(t)$ vectors of length $10^7$ stored as doubles and disregards all other factors. From our experience, for some $N_r \times N_t = 10^5$ threads at $\sim 2$ GHz the optimization would take an order of magnitude of 30 minutes of walltime to converge after about a hundred of iterations.

Currently, these requirements can be barely satisfied by booking a complete supercomputer such as JUROPA,[2] and any substantial increase in the number of units, or in the amount of data to be processed will put the problem beyond our reach. However, while the number of parameters of the complete log-likelihood function $\mathcal{L}$ in our formulation is $\mathcal{O}(\theta) \sim N^2$, thanks to the above-mentioned separability property, the number of parameters of $\mathcal{L}_i$ is linear in the number of units, $\mathcal{O}(\theta_i) \sim N$. Not only does this present major practical advantages such as easier scheduling of smaller jobs, but it also makes it possible to solve larger problems at all by proportionally trading the execution time for the amount of resources allocated to the optimization process.

### 2.6.3 Technical realization

Our model was implemented in Python, an increasingly popular language in the field of computational neuroscience. It relies upon the NumPy and SciPy scientific libraries[3] for essential data structures and algorithms. We used Cython[4] in order to bind to the OpenMP-parallelized computational kernels, that we extracted and re-wrote in C++ for performance reasons, and in order to access the mathematical functions from Boost C++ library. The distribution was implemented using the Python bindings to MPI, mpi4py.[5]

The optimization was performed via the NLopt[6] package by Steven G. Johnson using the low-storage Broyden-Fletcher-Goldfarb-Shanno method (Liu and Nocedal 1989) with support for bound constraints (Byrd et al. 1995) implemented by Ladislav Luksan (L-BFGS-B). We chose to use BFGS instead of the nonlinear conjugate gradient (CG) algorithm, because the former approximates the inverse Hessian matrix of the problem and uses it to steer the search in the parameter space. This results in improved convergence at the cost of higher iteration overhead. Since in our case the computation of the objective function is substantially more expensive, this trade-off is worthwhile.

As a stopping condition, we used a criterion based on the fractional tolerance of the objective function value. The optimization was terminated if $\eta = |\Delta \mathcal{L}| / |\mathcal{L}|$, where $\Delta \mathcal{L}$ is the decrease in the function value from one iteration to next, reached the threshold of $\tilde{\eta}$. The value of $\tilde{\eta}$ was selected close to the machine epsilon for the double precision floating point type, as requesting even lower tolerance would not yield a more accurate solution; the typical choice was $\tilde{\eta} \leq 10^{-15}$.

It is worth to note that in the case of $\ell_1$ regularized optimizations, it turned out that all gradient-based algorithms we tried were very much affected by the non-smoothness at zero, introduced by the regularization term in Eq. (16). A thorough review of the existing approaches to address this issue is presented in (Schmidt et al. 2009); we opted for implementing a smooth $\epsilon$-$\ell_1$ approximation, originally suggested in Lee et al. (2006):

$$\alpha \sum_{\substack{j=1 \\ j \neq i}}^{N} |J_{ij}| = \alpha \sum_{\substack{j=1 \\ j \neq i}}^{N} \sqrt{J_{ij}^2 + \epsilon}, \text{ for } \epsilon \to 0. \quad (17)$$

The derivatives of $\mathcal{L}_i$ with respect to $J_{ij}$ (12) have to be adjusted by addition of $-\alpha J_{ij} / \sqrt{J_{ij}^2 + \epsilon}$ respectively. We found that this approximation works well in practice for sufficiently small values of $\epsilon < 10^{-7}$ and enables us to use the L-BFGS-B algorithm without modifications. Additionally, we imposed bound constraints on the model parameters as discussed in Section 2.5; typical constraint ranges were $|J_{ij}| < 50$ mV for synaptic weights and $0.001$ s$^{-1} < c_i < 100$ s$^{-1}$ for base rates. The recordings were truncated to the first and last recorded spikes, $T_0 = t_1$ and $T_1 = t_q$, where $q$ is the total number of recorded spikes.

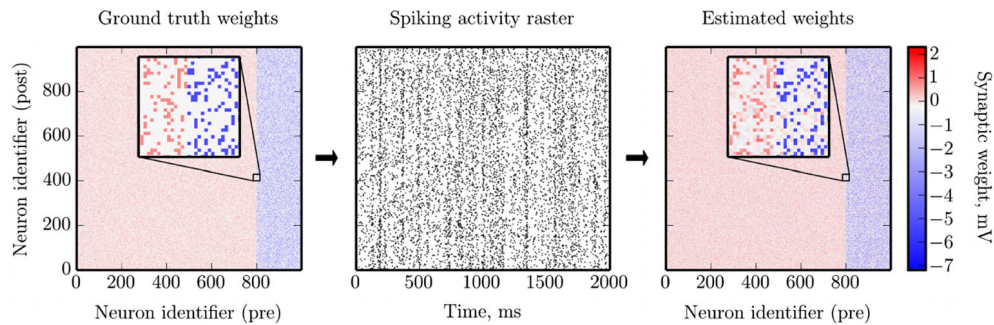

**Fig. 2** Schematic representation of the method validation setting. A test neural network is set up using known ground truth connectivity matrix (*left panel*) and its dynamics is numerically simulated. The emerging spiking activity of the neurons (*middle panel*) is recorded and fed into the reconstruction procedure. The resulting connectivity matrix (*right panel*) is then compared to the original one to assess the performance of the proposed technique. Insets show a zoom-in of the connectivity matrix, as indicated

# 3 Results

We quantified the effectiveness of our suggested method by performing a series of experiments as illustrated in Fig. 2. In these experiments we simulated neuronal networks with known (ground truth) connectivity, and reconstructed the synaptic weight matrix along with the model parameters of these networks on the basis of the recorded spike times. In this way, estimation results could be readily compared to the original connectivity matrix and model parameters. All simulations presented in this section were carried out with the NEural Simulation Tool (NEST) (Gewaltig and Diesmann 2007) and reconstructions were performed using the CPU implementation of the MLE optimizer as described in Section 2. Although the connectivity is sparse in all experiments considered below, we generally use MLE optimization here; only in Section 3.3, which describes the most difficult of the experiments, we also use regularization in order to demonstrate that our computational framework can handle regularized optimization.

In the following subsections, we present the benchmarks of the proposed technique against simulations of a widely used model of a random balanced network (Brunel 2000) and investigate the effect of choosing different neuron and synapse models, first with homogeneous and then with randomly distributed parameters. Finally, we show a successful reconstruction of a specific, non-random network, a "synfire chain" embedded in a balanced random network, only from "background" network activity (where the chain was not stimulated). Finally, by stimulating the synfire chain in a simulation of the estimated model, and comparing resulting dynamics to the output of the original network, we highlight the generative aspect of GLM network models. The abbreviations used in the following sections are summarized in Table 1.

## 3.1 Random balanced network of GLM neurons

As an initial testbed for our method, we selected a random balanced neural network of excitatory and inhibitory neurons in the asynchronous irregular (AI) spiking regime (Brunel 2000). Random networks do not have any particular structural features that can be exploited by the optimizer in order to improve the quality of the reconstruction, and hence in this sense they represent a "worst-case" type of input that is particularly useful for benchmarking purposes. Such networks are commonly studied using the leaky integrate-and-fire (LIF) neuron model. However, in order to be able to interpret the follow-up experiments, we first chose to assess the performance of our estimation method under idealized conditions, in which the simulated and estimated neuron and synapse models coincide: the GLM neuron model as described in Section 2 and simple synapses with exponential post-synaptic potentials.

**Table 1** Glossary of abbreviations

| | |
| --- | --- |
| EM | expectation-maximization |
| GLM | generalized linear model |
| GMM | Gaussian mixture model |
| KDE | kernel density estimation |
| LIF | leaky integrate-and-fire |
| MER | misclassification error rate |
| MLE | maximum likelihood estimation |
| PDF | probability density function |

As discussed in Section 2.2, given several conditions that our GLM satisfies, there is a unique maximum likelihood parameter set for the estimated network model (Paninski 2004). In the limit of an infinite amount of spike data used for model estimation and arbitrarily precise calculations, our method is thus bound to recover the true parameters of the simulated model. Hence, testing the method under idealized conditions, but for finite datasets, allows us to distinguish errors that are purely due to the limited length of the observations and restricted machine precision, from those due to a mismatch between the dynamics of the neuron and synapse models used to generate the data, and the dynamics of the models used to reconstruct the network.

The test network consisted of $N = 1000$ GLM neurons with 80 % : 20 % proportion of excitatory to inhibitory neurons ("pp_psc_delta" model in NEST nomenclature, with a base rate $c = 5\,\mathrm{s}^{-1}$, membrane time constant of $\tau = 20$ ms and a resting potential of $V_\mathrm{r} = 0$ mV). The nonlinearity gain of the neurons was set to $\delta u = 4$ mV as in Jolivet et al. (2006), which defines the scaling and units of a single post-synaptic potential via Eq. (5) ($\delta u = 1$ is assumed previously in Section 2 for the sake of convenience would make it unitless). Each connection was realized independently with a connection probability of $\epsilon = 0.2$ (Erdős-Rényi p-graph). The neurons were connected by synapses with exponential post-synaptic potentials with a peak amplitude of $J_\mathrm{e} = 1$ mV for excitatory and $J_\mathrm{i} = -5$ mV for inhibitory synapses, and a transmission delay of $d = 1.5$ ms. A strong inhibitory self-connection with $J_\mathrm{s} = -25$ mV and a transmission delay of $d_\mathrm{s} = \Delta t$ was used to model post-spike effects. The simulation progressed in time steps of $\Delta t = 0.1$ ms (resolution) and the simulation time was $T = 1$ hour. The average firing rate of the neurons was $\nu = 4.2\,\mathrm{s}^{-1}$. The recorded spike trains were fed to the estimation method, assuming known values of the time constant $\tau$, the transmission delays $d$ and the delay of the self-connection $d_\mathrm{s}$. The method produced estimates of the synaptic weight matrix $J_{ij}$ and the base rates $\{c_i\}$ for all neurons. The original and reconstructed synaptic weight matrix for this experiment are presented in Fig. 2. Throughout this text we refer to $\{J_{ij}\}_{1 \le i, j \le N,\ i \ne j}$ as the weight matrix; the self-connections $\{J_{ii}\}_{1 \le i \le N}$ and the baseline potentials $\{J_{i0}\}_{1 \le i \le N}$ are treated separately.

In order to evaluate the quality of the reconstruction, we analyzed the resulting distributions of recovered synaptic weights and base rates, as shown in Fig. 3. Whereas the probability density function (PDF) of the original distribution of synaptic weights can be described as a sum of three $\delta$-functions (for excitatory, inhibitory and null connections respectively), the peaks in the reconstructed distribution are broader due to the finite duration of the recording and limited machine precision, to the extent that for realistic values of parameters, there is a degree of overlap between the components of the distributions that represent excitatory and null connections. We noted that the amplitude of the noise that causes the broadening decreases approximately

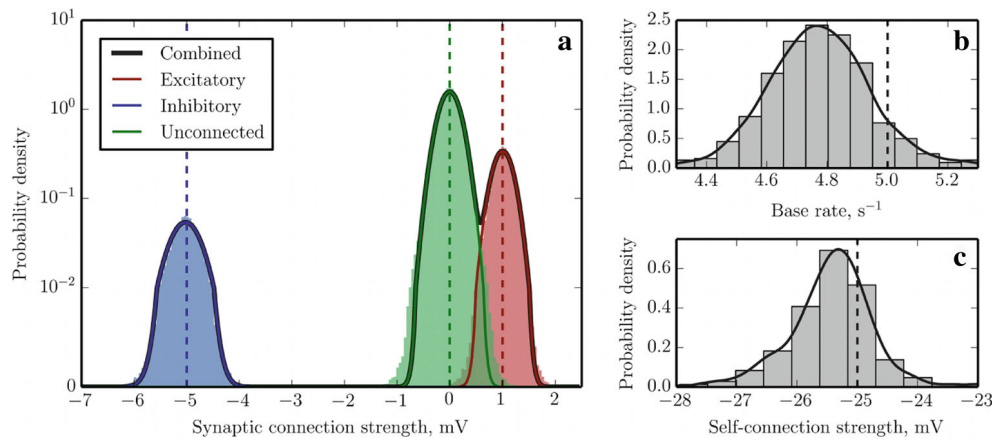

**Fig. 3** Reconstruction of a random balanced network of GLM neurons. The reconstruction was performed for $\tau = 20$ ms, $d = 1.5$ ms and $d_\mathrm{s} = 0.1$ ms. **a** Gaussian Mixture Model fit for the probability density function of the elements of the reconstructed synaptic weight matrix $J$ (*black solid curve*) and individual components contributed by excitatory (*red solid curve*, $\langle J_\mathrm{e} \rangle = 1.004$ mV), inhibitory (*blue solid curve*, $\langle J_\mathrm{i} \rangle = -5.023$ mV) and null (*green solid curve*, $\langle J_\varnothing \rangle = -0.002$ mV) connections. For comparison, we plot as histograms of $n = 200$ bins the distributions of the reconstructed synaptic weights, partitioned into three classes and colored according to whether the corresponding entry in the ground truth connectivity

matrix was $J_\varnothing = 0$ mV (unconnected; *green*), $J_\mathrm{e} = 1$ mV (excitatory; *red*) or $J_\mathrm{i} = -5$ mV (inhibitory; *blue*). A perfect reconstruction would result in delta peaks at the three synaptic strength values of the original connectivity matrix, marked with *red, blue and green dashed vertical lines*. The scale of the vertical axis is logarithmic, except for the first decade, which is in linear scale. **b, c** Distributions of the identified base rates of the neurons and weights of the self-connections approximated with histograms and Gaussian KDEs. *Black dashed vertical lines* mark the ground truth values which should have been recovered ($c = 5\,\mathrm{s}^{-1}$ and $J_\mathrm{s} = -25$ mV respectively) in inverse proportion to the square root of the duration of the recording (data not shown), however, we selected $T = 1$ hour as a reasonable standard amount of input data to mimic conditions where the duration of the recording is limited due to experimental and computational constraints.

This circumstance thus makes it difficult to identify weak excitatory connections unambiguously, and therefore an advanced approach to classification was needed to obtain optimal network reconstruction. To this end, we fitted a Gaussian mixture model (GMM) with a fixed number of components ($n = 3$) to the reconstructed synaptic weights, assuming that synaptic connections, in general, can be either excitatory or inhibitory, or absent. We used an expectation-maximization (EM) algorithm to obtain a maximum likelihood estimate (MLE) of the GMM parameters (mixing weights, means and variances of the individual components), and classified the synaptic weights accordingly. The fitting and classification was performed using a Python implementation of GMM (sklearn.mixture.GMM) provided by the scikit-learn toolkit (Pedregosa et al. 2011). In order to reconstruct the PDFs of the base rates and self-connections, we used both the histogram function from the NumPy library and the Gaussian kernel density estimation (KDE) code from the SciPy library.

The results are illustrated in Fig. 3, which shows that the means of the distributions were almost perfectly reconstructed and that GMM is indeed an appropriate model for this PDF. The recovered base rates and self-connection weights are also more or less in agreement with the ground truth values. The detailed classification performance breakdown is presented in Table 2, showing that the classification of synaptic connections is nearly optimal for this dataset (assuming that the cost of making a "false positive" error is equal to the cost of the "false negative" error) and the number of misclassified connections is less than 1%.

**Table 2** Breakdown of classification errors for the GLM random network

| Connection type | Errors | FP | FN | ND |
|---|---|---|---|---|
| Excitatory | 7 300 | 62 % | 38 % | 15 % |
| Inhibitory | 0 | | | |
| Unconnected | 7 300 | 38 % | 62 % | — |
| Total errors | 0.73 % | | | |

The column "Errors" shows the absolute number of incorrectly classified connections belonging to each class. The columns "FP" and "FN" show the percentage of false positives and false negatives of this number accordingly. The column "ND" is the percentage of misclassified connections in violation of the Dale's law (i.e. an inhibitory neuron is assigned an outgoing excitatory connection, or vice versa), which was not enforced for this reconstruction. The last row shows the percentage of erroneously classified connections of the total number of possible connections ($N^2 = 10^6$ for $N = 1000$)

## 3.2 Random balanced network of LIF neurons

Having established the baseline performance in ideal conditions, we designed our next experiment to gauge the influence of mismatch between the neuron and synapse models used to generate the data and those used to reconstruct the network. To this end, we generated data with the commonly used, more complex and realistic LIF neuron model with $\alpha$-shaped post-synaptic currents (PSCs). We then carried out the reconstruction as before assuming our simplified GLM neuron model and synapses with exponential post-synaptic potentials. Another important point is that whereas in the previous experiment we assumed that the membrane time constant $\tau$ and transmission delays between the neurons $d$ are known in advance, this is certainly not the case in the laboratory setting, and hence a principled way of estimating these parameters is required in order to analyze real physiological data.

To generate the test data, we wired a network similar to the one described in the previous section, but using a LIF instead of a GLM neuron model. As before, we used $N = 1000$ neurons with 80 % : 20 % ratio of excitatory to inhibitory cells, connection probability of $\epsilon = 0.2$ (each connection was realized independently), transmission delay of $d = 1.5$ ms, simulation resolution of $\Delta t = 0.1$ ms. Synaptic weights were set to $\hat{J}_{e/i} = J_{e/i} \times w$, with $J_e = 1$ mV and $J_i = -5$ mV. The latter ($J_e$ and $J_i$) were again interpreted as peak PSP amplitudes, where $w = w(\tau_m, \tau_s, C)$ was the scaling factor (specific to the post-synaptic neuron) selected such that an incoming spike passing through a connection with the synaptic weight of $w$ would evoke a PSP with the maximum amplitude of 1 mV. The parameters of the LIF model ("iaf_psc_alpha" in NEST nomenclature) were chosen as follows: membrane capacitance $C = 250$ pF, membrane time constant $\tau_m = 20$ ms, synaptic time constant $\tau_s = 0.5$ ms, refractory time $t_r = 2$ ms, firing threshold $\theta = 20$ mV, resting potential $V_r = 0$ mV and reset to $V_r$ after each spike. This time, additional to the synaptic input from other simulated neurons, each neuron received independent Poisson process excitatory inputs at a rate of $\nu_e = 1779$ s$^{-1}$ and inhibitory inputs at $\nu_i = 0.2 \times \nu_e = 356$ s$^{-1}$. These external inputs represent the influence of neurons that are not part of the simulation, and are necessary to achieve asynchronous and irregular activity as in cortical networks (Brunel 2000). The simulation time was set to $T = 2$ hours and the data was cut into training and validation parts of $T_t = T_v = 1$ hour as explained below. The average neuron firing rate was $\nu = 4.2$ s$^{-1}$, and so matched the average neuron firing rate of the network of the GLM neurons presented above.

In order to recover the GLM parameters $\tau$ and $d$ for this experiment, we applied a cross-validation procedure.

It is important to note that we are not expecting to obtain exactly $\tau = \tau_m = 20$ ms and $d = 1.5$ ms due to mismatch between the LIF with $\alpha$-shaped PSCs and GLM with exponential PSPs models. Instead, we want to recover the optimal parameters $\tau$ and $d$ for the GLM model to produce most similar dynamics to the recorded spike trains from the LIF model. We split the available data into a training and a validation dataset, and performed reconstructions for a subset of $N_s = 75$ neurons on the training dataset varying one parameter, while keeping the other one fixed. The resulting parameter estimates $\theta_i$ were then used to calculate the log-likelihood function $\mathcal{L}_i$ on the validation dataset. Two datasets (training and validation) were used in order to ensure that the chosen values of the parameters generalize, and are not specific to the training sample. The validation curves are shown in Fig. 4a, c (the curves for the training dataset look identical); note that they all have an easily identifiable maximum. Subsequently, we averaged the locations of the maxima for all trials and performed another cross-validation run (Fig. 4d, b) for updated values of the parameters. Repeating this procedure of alternatively fixing one parameter and performing cross-validation for another one would lead us to a local extremum in the $(\tau, d)$ parameter space. However, we opted to stop after only a few iterations because the procedure is computationally expensive, and in order to asses if a sub-optimal choice of $\tau = 10$ ms and $d = 1.7$ ms would lead to acceptable estimation results.

After determining $\tau = 10$ ms and $d = 1.7$ ms through the cross-validation procedure, we used these values to estimate the connectivity and base rates. The results of the connectivity reconstruction on the training dataset were processed in the same way as in the previous subsection and are presented in Fig. 5, with further details on the classification of synaptic connections in Table 3. We find that the reconstruction quality as defined by classification into the groups of excitatory, inhibitory and null connections closely matches the performance on the ideal dataset analyzed in the previous section, despite the mismatch in models and the suboptimal choice of $\tau$ and $d$. Note that in this experiment, the recovered values of synaptic weights in mV cannot be compared directly to the ones that were used in the simulation which produced the data due to the differences between GLM and LIF models, unlike in the first experiment described in Section 3.1. However, this does not matter for the purposes of classification.

### 3.3 Random balanced network with distributed parameters

To make the reconstruction task more challenging and to create a more realistic benchmark for our method, we amended the network described in the previous subsection to have different parameters $J_e$, $J_i$, $d$, $\tau_m$ and $\tau_s$ for every neuron and synaptic connection, sampled from uniform distributions around each respective mean value (Table 4),

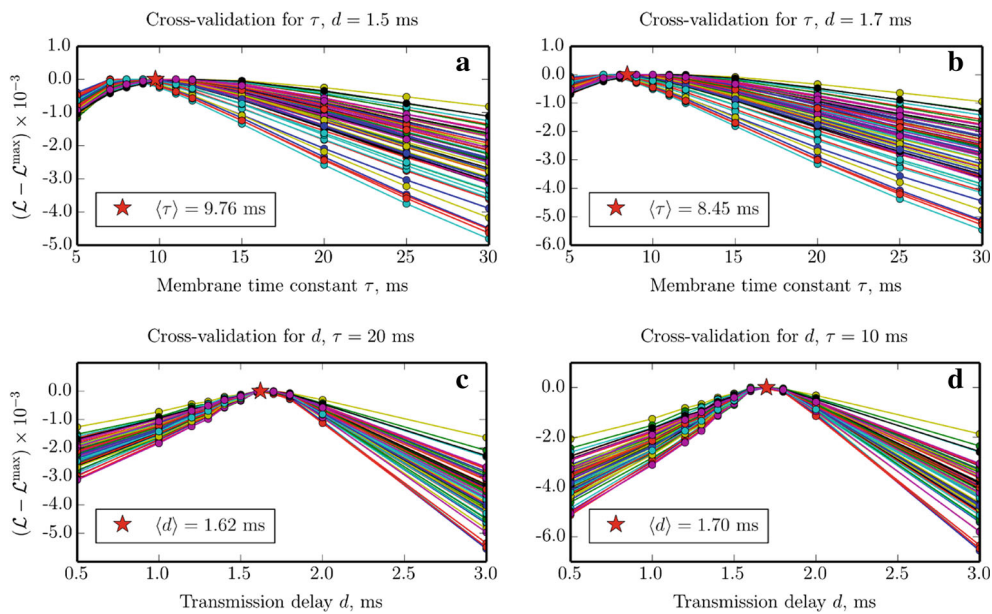

**Fig. 4** Cross-validation for the membrane time constant $\tau$ and transmission delay $d$. Log-likelihood $\mathcal{L}$ computed on the validation dataset, using parameters estimated from the training dataset for different values of parameters $\tau$ and $d$. For each trial, $\mathcal{L}$ has been rescaled according to $\mathcal{L} \leftarrow (\mathcal{L} - \max \mathcal{L}) \times 10^{-3}$. The *red star* marks the average of the horizontal location of the peaks of all *curves* in the plot.

Each panel shows $N_s = 75$ parameter scans for $N_e = 60$ excitatory and $N_i = 15$ inhibitory neurons, randomly selected from the complete recording of $N = 1000$ neurons. **a, b** Cross-validation for $\tau$ using fixed values of $d = 1.5$ ms (standard deviation $\sigma = 1.14$ ms) and $d = 1.7$ ms ($\sigma = 1.02$ ms). **c, d** Cross-validation for $d$ using fixed values of $\tau = 20$ ms ($\sigma = 0.04$ms) and $\tau = 10$ ms ($\sigma = 0.01$ ms)

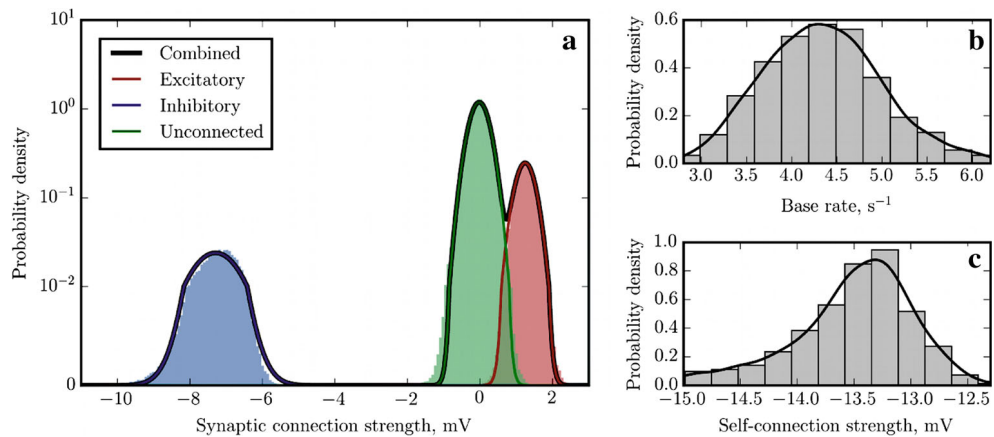

**Fig. 5** Reconstruction of a random balanced network of LIF neurons with α-shaped PSCs. The reconstruction was performed for $\tau = 10$ ms, $d = 1.7$ ms (obtained through cross-validation) and $d_s = 0.1$ ms. **a** GMM fit for the PDF of the reconstructed synaptic weight matrix (*black solid curve*) and individual components (colored *solid lines*); colored *bars* under the curves show the distributions of the reconstructed synaptic weights classified using the ground truth synaptic connectivity matrix as in Fig. 3, approximated as histograms of $n = 200$ bins. The scale of the vertical axis is logarithmic, except for the first decade, which is in linear scale. **b, c** Histograms and Gaussian KDEs approximating the PDFs of the base rates and self-connection weights

which are the same as in the previous experiment. However, instead of trying to recover the individual values of $\tau_i$ for each neuron and $d_i$ for every connection, we decided to investigate whether it would be still possible to make a useful reconstruction assuming identical "mean" values of $\tau$ for all neurons and $d$ for all connections. Additional motivation for this choice is in that cross-validation is a computationally expensive procedure: whereas individual estimation might converge in a matter of minutes, the amount of resources needed to scan a multidimensional parameter grid grows quickly and becomes unmanageable. Therefore, we performed cross-validation on a subset of neurons as described in the previous subsection, and settled for $\tau = 10$ ms and $d = 1.7$ ms again (data not shown).

The estimation results for this dataset are shown in Fig. 6 and Table 5 (left panel and left part of the table respectively). The PDFs of the reconstructed synaptic weights were approximated using Gaussian KDE. Obviously, the individual components of the PDF were distorted, because instead of using optimal values for $\tau_i$ and $d_i$, we used rather arbitrarily chosen fixed values for all neurons and connections. However, more importantly, as the components of the original PDF of synaptic weights were broad distributions rather than δ-functions, the resulting recovered distribution components are strongly non-Gaussian. Therefore, in this case the EM procedure for GMM fails to converge to reasonable means and variances, and is no longer a viable choice to perform the classification of connections.

However, instead of engaging in more elaborate statistical modeling to overcome this difficulty, we can take a step back and resort to an unsupervised learning technique called *k*-means clustering (which is actually a simplification of GMM). This method rejects the probabilistic assignment of data points to components, and instead makes the assumption that each point belongs to one (and only one) cluster, to the centroid of which it is closest in terms of Euclidean distance. This simplification leads to sub-optimal classification

**Table 3** Breakdown of classification errors for the LIF random network with α-shaped PSCs

| Connection type | Errors | FP | FN | ND |
|---|---|---|---|---|
| Excitatory | 7 020 | 58 % | 42 % | 10 % |
| Inhibitory | 2 | 100 % | 0 % | 100 % |
| Unconnected | 7 022 | 42 % | 58 % | — |
| Total errors | 0.70 % | | | |

The meaning of the abbreviations is the same as in Table 2

**Table 4** Distribution of parameter values of a random balanced network of LIF neurons

| Parameter | Symbol | Range | Spread |
|---|---|---|---|
| Excitatory weight | $J_e$ | 0.8 … 1.2 mV | ±20 % |
| Inhibitory weight | $J_i$ | −4 … − 6 mV | ±20 % |
| Transmission delay | $d$ | 1 … 2 ms | ±33 % |
| Membrane time constant | $\tau_m$ | 15 … 25 ms | ±25 % |
| Synaptic time constant | $\tau_s$ | 0.3 … 0.7 ms | ±40 % |

All parameters except for $d$ were sampled from continuous uniform distributions limited by the values in the table. The transmission delay $d$ was sampled from a discrete distribution with the step equal to the simulation resolution $\Delta t = 0.1$ ms. Note that the synaptic scaling factor $w$ depends on $\tau_m$ and $\tau_s$ of the post-synaptic neuron, and, therefore, the distributions for $\hat{J}_e$ and $\hat{J}_i$ ("ground truth weights") were not uniform

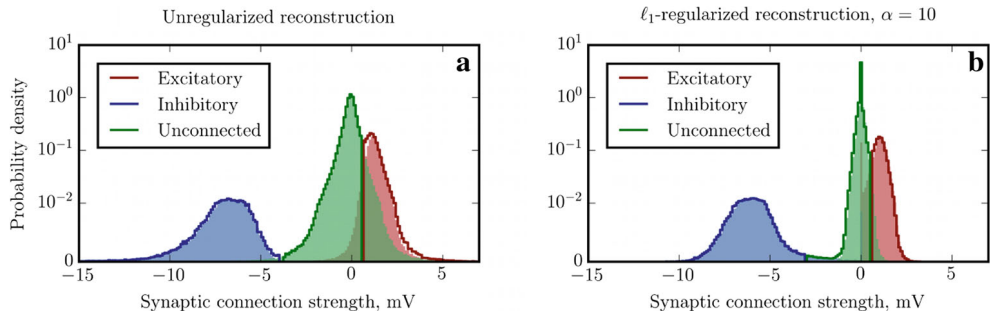

**Fig. 6** Unregularized and $\ell_1$-regularized reconstruction of a random balanced network of LIF neurons with $\alpha$-shaped PSCs and distributed parameters. The colored *solid step lines* show PDFs approximated with histograms of $n = 200$ bins of the reconstructed synaptic weights corresponding to the classification via $k$-means clustering. *Vertical lines* demarking the boundaries between distributions designate the points that are equidistant from the identified centroids. The colored *bars* under the *curves* represent PDFs estimated by histograms of $n = 200$ bins, classified using the ground truth connectivity matrix, as in Figs. 3 and 5

when the underlying distributions violate these constraints, but the resulting algorithm is fast and robust.

The Voronoi diagrams for $k$-means classification are represented in Fig. 6 as solid lines: the colors show which of the three centroids is closest, in blue, green and red for inhibitory, null and excitatory connections, respectively. By comparing the solid curves and envelope of the colored bars it can be seen that in this case there is a significant overlap between the components contributed by null connections and excitatory connections. Therefore, even the most advanced classification strategies will lead to a substantially higher amount of classification errors than in the previous experiments. The classification data using $k$-means is given in Table 5 (left part).

Nevertheless, the situation can still be considerably improved: here, we exploited the sparsity of the synaptic connection matrix by regularizing the GLM estimation with a $\ell_1$ penalty term as explained in Section 2.5. Imposing such a prior on the estimation causes shrinking of the distribution of null connections (Tibshirani 1996) and thus enables better separation between the components. However, the choice of the penalty scaling constant $\alpha$ is arbitrary and so we again availed ourselves of a cross-validation procedure to determine the optimal value for our dataset.

The results of the reconstruction for a subset of the recorded neurons with different values of $\alpha$ on the training dataset are shown in the left panel of Fig. 7. The right panel depicts the subsequent evaluation of the log-likelihood function on the validation dataset. It is important to note that, for optimal results, this procedure should generally be performed for all neurons, and an individual regularization coefficient should be selected for each of the cells. Instead, in order to save computational resources, we only performed it for a subpopulation of neurons and subsequently selected the same value of $\alpha = 10$ for all cells, which is slightly lower than the average, to prevent excessive connection pruning in neurons with small optimal $\alpha$.

We performed a full $\ell_1$-regularized GLM estimation using $\alpha = 10$, still fixing the parameters to $\tau = 10$ ms and $d = 1.7$ ms, the results of which are presented in Fig. 6, right panel and Table 5, right part. The plot shows that the contribution by null connections indeed shrunk significantly, and thus the amount of classification errors was decreased almost by half. At the same time, for some neurons $\alpha = 10$ turned out to be too strong of a regularization factor, and thus the estimator, in an overzealous attempt to find a sparse solution, set to zero some of the weaker excitatory and inhibitory synapse weights. This can be seen as

**Table 5** Breakdown of the classification errors for the unregularized and $\ell_1$-regularized ($\alpha = 10$) connectivity estimations of a LIF random network with distributed parameters

| Connection type | Unregularized | | | | Regularized, $\alpha = 10$ | | | |
|---|---|---|---|---|---|---|---|---|
| | Errors | FP | FN | ND | Errors | FP | FN | ND |
| Excitatory | 57 827 | 84 % | 16 % | 17 % | 33 706 | 16 % | 84 % | 3 % |
| Inhibitory | 2 226 | 95 % | 5 % | 79 % | 2 673 | 0 % | 100 % | 0 % |
| Unconnected | 59 749 | 16 % | 84 % | — | 36 379 | 85 % | 15 % | — |
| Total errors | 5.99 % | | | | 3.64 % | | | |

The categories of errors are the same as in Table 2

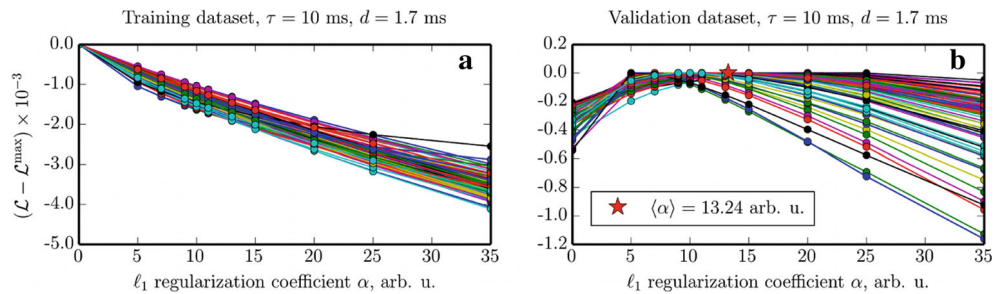

**Fig. 7** Cross-validation for the $\ell_1$ regularization coefficient $\alpha$ for fixed values of $\tau = 10$ ms and $d = 1.7$ ms. Each panel shows $N_s = 75$ parameter scans for $N_e = 60$ excitatory and $N_i = 15$ inhibitory neurons randomly selected from the complete recording of $N = 1000$ neurons. **a** The values of the rescaled log-likelihood $\mathcal{L}$ as a function of the $\ell_1$ regularization coefficient $\alpha$ computed on the training dataset. **b** Log-likelihood $\mathcal{L}$ as a function of $\alpha$ computed on the validation dataset using the parameters estimated from the training dataset. The *red star* marks the average of the horizontal location of the peaks of all *curves* in the plot

a secondary peak of the red distribution at the origin. A secondary peak of the blue distribution is also present, but scarcely visible due to scale.

### 3.4 Synfire chain embedded in a random balanced network

#### 3.4.1 Construction of the network model

In this experiment, we turned to structured networks in order to highlight the generative aspects of the proposed GLM model and demonstrate a potential approach to the interpretation of the recovered connectivity. One specific structure of interest, prominent in the context of cortical networks, is called a "synfire chain" (Abeles 1982). The synfire chain, consisting of consecutively linked and synchronously activated groups of neurons, is a thoroughly studied model of signal propagation in the cortex (Diesmann et al. 1999; Goedeke and Diesmann 2008).

We built a simulation of a random balanced network with an embedded synfire chain, simulated the dynamics of this network and recorded its spiking activity, which we then used as input data for the MLE procedure to infer the parameters of our GLM (no regularization was applied in this experiment, unlike in the last case presented in Section 3.3). However, as would be the case with the experimental recordings, we did not assume that we know the "right" ordering of the neuron identifiers. We therefore subjected the recovered connectivity to a clustering process in order to reveal the trace of the synfire chain in the connection matrix. After identifying the synfire chain in the network, we performed a simulation where we stimulated the discovered first "link" of the chain in the original and reconstructed networks, and observed identical dynamics in both cases.

Similarly to the previous experiments, we first constructed a random balanced network of LIF neurons ($N = 1000$) with 80 % : 20 % proportion of excitatory to inhibitory cells. This time, we used "iaf_psc_delta_canon" model in NEST nomenclature; this model is different from the standard "iaf_psc_delta" and "iaf_psc_alpha" LIF neurons in that the points in time when it emits spikes are not tied to the grid defined by the simulation resolution, but rather are recorded precisely as they occur (Morrison et al. 2007; Hanuschkin et al. 2010). Correspondingly, for the external inputs, we employed the continuous time version of the Poisson generator "poisson_generator_ps". Since this network model works in continuous time and does not require discretization or binning of the spike data, we wanted to examine the implications of feeding the precise spike times to the MLE of the GLM, as opposed to data binned to $\Delta t = 0.1$ ms simulation resolution as in the previous experiments. The model parameters were fixed to $\tau_m = 20$ ms, $\tau_r = 2$ ms, $\theta = 20$ mV, and $V_r = 0$ mV. Each neuron was set to receive a fixed number of incoming connections ($M_e = 80$ excitatory and $M_i = 20$ inhibitory), where the pre-synaptic neurons were randomly selected (without replacement) from the excitatory and inhibitory populations respectively (implemented as "RandomConvergentConnect" function in NEST). Synaptic weights were set to $J_e = 0.9$ mV for excitatory, $J_i = -4.5$ mV for inhibitory connections with a transmission delay of $d = 1.5$ ms. Additional independent Poisson process excitatory inputs were supplied at $\nu_e = 2222$ s$^{-1}$ and inhibitory inputs at $\nu_i = 0.25 \times \nu_e = 556$ s$^{-1}$.

On top of this "background" network, we selected $N_l = 10$ groups (links) of $N^\alpha = 50$ neurons each ($N_e^\alpha = 40$ excitatory and $N_i^\alpha = 10$ inhibitory cells) and connected all $N_e^\alpha$ excitatory neurons of every group to each of the $N^\alpha$ neurons in the next group with $J_e^\alpha = 1.4$ mV excitatory synapses (transmission delay $d = 1.5$ ms). Inhibitory neurons in a link of the chain do not have specific connections to the next link in the chain (Hayon et al. 2004). No neuron in the network was part of more than one group of the synfire chain. This way, we created a "hidden" embedded synfire chain, which receives inputs from the background random network and likewise projects outgoing connections to the background network. When the first group of this structure is stimulated in a coordinated fashion, the chain reliably propagates the excitation from one group to the next until it reaches the last one, and terminates. In the absence of such coordinated stimulation, the synfire chain did not activate, and only "background" activity was observed.

### 3.4.2 Identification of the synfire chain by connectivity clustering

The complete network was simulated for $T = 2$ hours of biological time and exhibited an average firing rate of $\nu = 1.4\,\mathrm{s}^{-1}$. The synfire chain was not stimulated during the simulation, so the spike train recordings contained no instances of propagating synfire activity. The neuron identifiers were randomly shuffled and the resulting spike raster was fed into the MLE reconstruction procedure.

We reasoned that one of the most generic differentiators between the neurons that belong to various groups (inhibitory neurons and excitatory neurons that are, or are not part of the synfire chain) is the relative strengths of the synapses (both incoming and outgoing connections can be considered). Therefore, we can apply a clustering algorithm to the recovered connectivity matrix to discern between several classes of neurons. However, most algorithms (such as $k$-means or GMM, employed in the previous sections) require the desired number of clusters to be set explicitly, either through prior knowledge, or by applying statistical or information theory methods to the data to get an estimation.

To circumvent this problem, we carried out an unsupervised learning technique known as hierarchical clustering. It amounts to iteratively repeating the procedure of looking at the discovered clusters (which, in the first step, each contain a single element), determining the ones that are most similar according to a chosen metric, and merging them into an agglomerate cluster; the process continues until a single cluster remains. The results are visualized by constructing a so-called "dendrogram", which shows the discovered hierarchy of clusters as a tree structure. Therefore, it is not necessary to specify the number of clusters in advance, but rather the most appropriate set of clusters can be selected by analyzing the dendrogram after performing the clustering. This approach fits very well to an exploratory setting, where one might wish to appreciate the entirety of possible groupings in a compact graphical form and then choose the one that best highlights the particular aspect of interest of the data.

We applied hierarchical clustering to the connectivity matrix using Ward's minimum variance method (Ward 1963) as a criterion for choosing the pair of clusters to merge at each step. Ward's minimum variance criterion minimizes the total within-cluster variance and enables the grouping of items into sets such that they are maximally similar to each other according to some definition of similarity, which is usually expressed in form of a "dissimilarity matrix". We used the SciPy hierarchical clustering package (scipy.cluster.hierarchy) to obtain the linkage and visualize the results.

Initially, we grouped the neurons by using the outgoing synaptic weights as the measure of dissimilarity, as shown for the MLE-reconstructed connectivity in Fig. 8a. This clustering enabled us to tell excitatory and inhibitory neurons apart (smaller blueish group on the left, and larger reddish group on the right of the matrix). Additionally, in this figure, we can see eight big red squares, which represent the links of the synfire chain. In total, nine squares should be visible in the connectivity matrix for $N_l = 10$ links, because the outgoing connections of the last link are not statistically different from those of the background neurons.

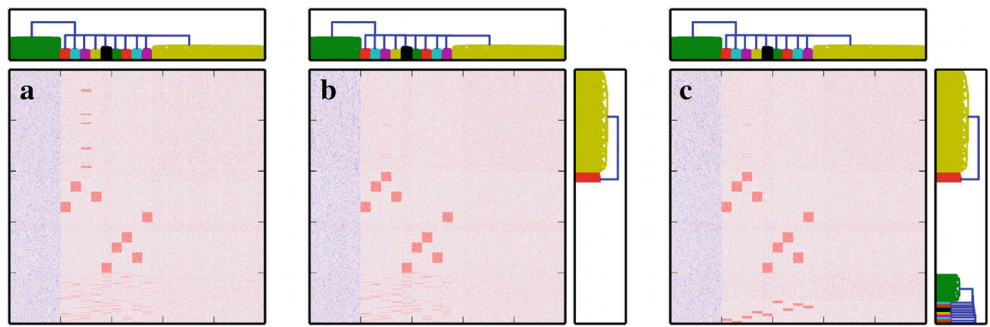

**Fig. 8** Clustering of synaptic weights uncovers the synfire chain in the reconstructed connectivity matrix. **a** Connectivity is first clustered by the outgoing connections (columns), while trying to achieve minimal variability inside each group. The dendrogram at the *top* of the panel shows the hierarchy of the clusters with the relevant groups highlighted in different colors. The *green cluster* on the *left* is formed by the inhibitory neurons. The *yellow cluster* at the *right* consists of excitatory neurons that are not part of the synfire chain and so do not have strong outgoing connections. The clusters in the middle correspond to the links of the synfire chain, which coalesce as *red squares* in the matrix. **b** Clustering by incoming connections (*rows*) inside the *yellow cluster* of neurons reveals the last link of the chain. **c** Clustering by incoming connections inside the *green cluster* helps to identify the inhibitory neurons that are part of the synfire chain (*thin red rectangles*)

The square missing from Fig. 8a is the last link of the chain, which by construction cannot be detected via clustering by the outgoing connections. Therefore, we subjected the neurons that are part of the big yellow cluster (excitatory neurons, which have not been previously identified as taking part in any of the synfire chain links) to additional clustering by incoming connections. This operation reveals the formerly concealed last link of the chain (Fig. 8b). Finally, we applied the same procedure to the inhibitory neurons in the big green cluster. This reveals the inhibitory neurons that are part of the synfire chain. These neurons receive connections from the previous link in the chain but do not send outgoing projections to the next links, and so they are also impossible to detect by clustering only by outgoing connections. This step completes the clustering procedure and we arrive at the final result as shown in Fig. 8c.

In Fig. 9, the clustered matrices (middle column) are contrasted with the matrices in randomized (left column) and

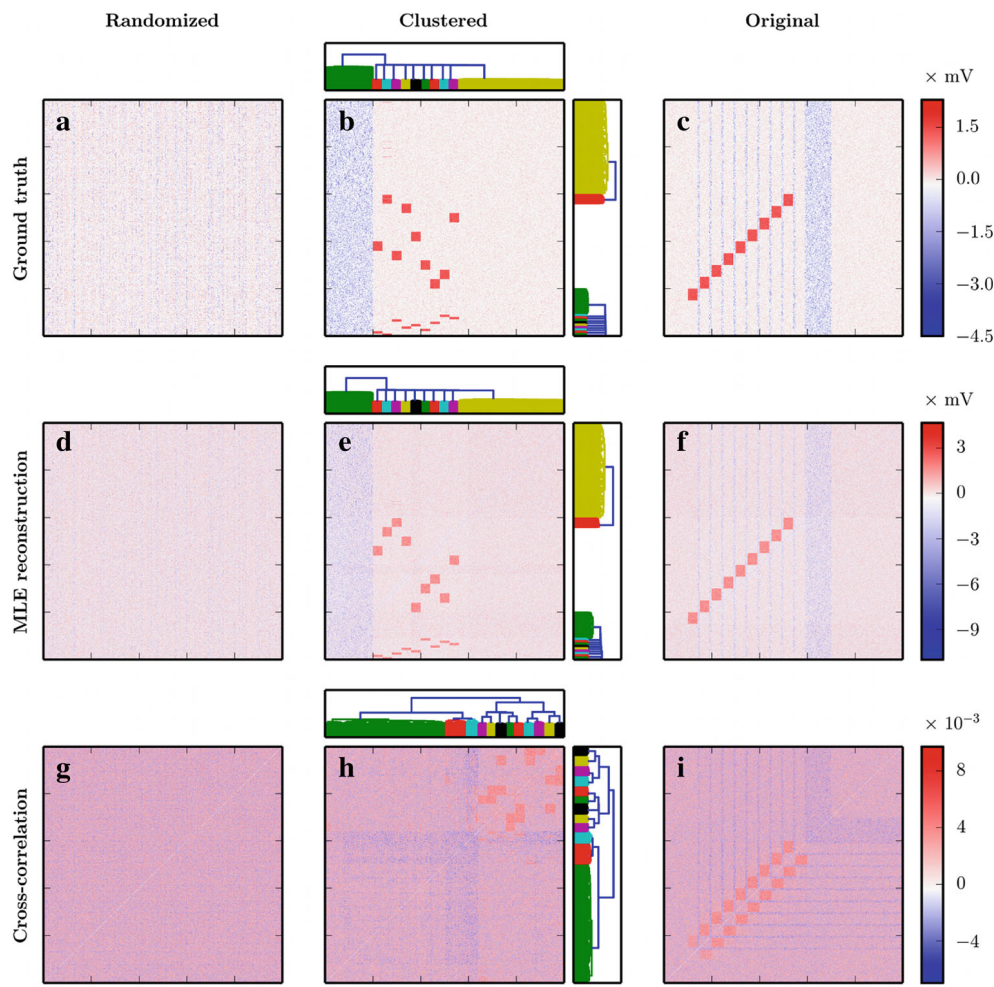

**Fig. 9** Identification of an embedded synfire chain by clustering connectivity estimated from "background" spiking activity. The grouping by rows delineates the panels produced on the basis of the ground truth connectivity, connections estimated using the GLM model, and lagged cross-correlation data. The grouping by columns lays out the panels presenting the connectivity matrices where the order of the neuron identifiers have been randomized, recovered by clustering and defined by the sequence in which the neurons were originally wired up. **a, d, g** Ground truth and MLE reconstructed synaptic weights, as well as lagged cross-correlation coefficient matrices for randomized neuron identifier order. **c, f** The *red rectangles* correspond to the connections from one chain link to the next. *Thin blue bands* identify inhibitory neurons that belong to the synfire chain. The *wide blue band* corresponds to the inhibitory neurons that are not part of the chain. **b,**

**e** The interpretation of the *bigger red rectangles* and the *wide blue band* is the same as above, except that all inhibitory neurons are now grouped together. The *thin red rectangles* at the *bottom* correspond to the groups of inhibitory neurons in the synfire chain receiving incoming connections from all excitatory neurons in the previous link. The clustering process that produced the reordering and the dendrograms is illustrated in steps in Fig. 8. **g, h, i** The lagged cross-correlation matrix is symmetric by construction. Therefore, the dendrograms at the *top* and on the *right* of panel h are identical (unlike **b, e**). Diagonal entries (all 1) were excluded here. **h, i** The *red rectangles* correspond to groups of neurons that exhibit positively correlated firing activity, while inhibitory neurons display negative correlation, marked by the *blue bands* original ordering (right column), i.e. the initial indexing of neurons that we used to define the neuron groups of the synfire chain network. An identical clustering procedure was applied to the ground truth connectivity matrix (Fig. 9a–c) and the one obtained from MLE estimation using the recorded spike trains (Fig. 9d–f). Note that, as explained at the end of Section 3.2, the reconstructed values of the synaptic weights in the second row cannot be directly compared to the original synaptic strengths.

The synfire chain is not apparent in the connectivity matrix in randomized ordering, neither for the ground truth matrix (Fig. 9a), nor the MLE-estimated connectivity (Fig. 9d). However, clustering neurons by the similarity of incoming and outgoing connection weights reveals the synfire chain substructure (Fig. 9b, e) of both excitatory and inhibitory neurons. This shows that our clustering procedure successfully recovers the group structure of the synfire chain network. Note that in the original ordering (Fig. 9f), the reconstructed matrix also resembles the ground truth matrix to a great extent (Fig. 9c), as expected based on our previous reconstruction experiments above.

### 3.4.3 Comparison to correlation-based connectivity estimation

In order to compare the results obtained using our GLM method with a well-established reference, we also performed lagged cross-correlation analysis on the same dataset. We computed the cross-correlation curves $\rho_{ij}(\tau)$ for all pairs of neurons with a bin size of $\Delta = 10$ ms and a maximum time lag of $\tau_{max} = \pm 200$ ms. The normalized Pearson cross-correlation coefficient for a stationary ergodic point process for sufficiently large number of sampled bins $K$ is defined as follows (Shao and Chen 1987):

$$\rho_{ij}(\tau) = \frac{\sum_{k=0}^{K} \tilde{S}_i(k\Delta) \tilde{S}_j(k\Delta + \tau) - N_i N_j K^{-1}}{\sqrt{\left(N_i - N_i^2 K^{-1}\right) \left(N_j - N_j^2 K^{-1}\right)}} .$$

Here, $\tilde{S}_i(t)$ and $\tilde{S}_j(t)$ are binned spike trains of neurons $i$ and $j$ (both $K$-bins long), whereas $N_i$ and $N_j$ are the total numbers of spikes of the respective neurons. For each $\rho(\tau)$ curve, we found the absolute extrema $\tau_{peak}^{ij} = \text{argmax}_\tau (|\rho_{ij}(\tau)|)$ and represented the results as a matrix of lagged cross-correlation coefficients $\tilde{J}_{ij} = \rho_{ij}(\tau_{peak}^{ij})$, shown in Fig. 9g–i.

We performed clustering on the $\tilde{J}_{ij}$ matrix as previously described, however, we had to limit ourselves to the first step only, because the cross-correlation matrix is symmetric by construction. The matrix shows similar patterns to the ground truth and MLE connectivity matrices, albeit with substantially lower contrast. Additionally, the direction of the synfire chain cannot be detected, due to the symmetry of the measure mentioned above. Moreover, the individual values of the correlation matrix are difficult to directly relate to the experimental quantities and/or model parameters, because the correlation matrix alone does not constitute a generative model, as we discuss below.

### 3.4.4 Simulation of original and reconstructed synfire chain networks

Finally, we compared the dynamics of the original and reconstructed network in simulation, including occasional stimulation of the first group of the synfire chain. In both networks, we can identify the order of the groups of the chain by following the links backwards starting from the last link identified in Fig. 8b. Note that the identification of the last link is not determined by the clustering algorithm but simply by membership of the neurons as pre- and post-synaptic partners in the strong connections represented as red boxes in the clustered connectivity matrix. Neurons in the last link occur just as post-synaptic targets; there is no red box in which they occur as pre-synaptic sources. Conversely the neurons of the first link only occur as pre-synaptic sources. Thus the chain can be unrolled from either end by analogous processes.

The simulation of the reconstructed GLM network was carried out in NEST using "pp_psc_delta" neurons and the recovered connectivity matrix. In order to avoid the necessity of fine-tuning the parameters of the stimulation, we additionally included a reset of the membrane potential $U_i(t) \leftarrow 0$ after spike emission (option "with_reset" in the "pp_psc_delta" model, enabled for all neurons), which prevents runaway excitation of the neurons in the network upon delivering a strong stimulus to the synfire chain.

The results of this experiment are displayed in Fig. 10. We used the order of the neuron identifiers in which the cells were originally wired up to permit a clear visualization of the activity. The raster plots show that the dynamics of the spike patterns of ground truth and estimated network are very similar. Generally, an estimate of a GLM based on recorded spike trains is a generative model of the data, in the sense that, if itself simulated, will produce similar data; Fig. 10 demonstrates this using our embedded synfire chain example.

## 4 Discussion

In the present work, we introduce a method for analysis of parallel spike trains based upon maximum likelihood estimation of parameters of a recurrent network of stochastic generalized linear model neurons. The method not only makes it possible to perform large-scale reconstruction of the directed synaptic connectivity of neuronal circuits, but also to recover neuronal parameters, which can be used to obtain a dynamic (i.e. simulatable) model of the network under investigation. Through radical simplification of the single neuron model and interaction kernels as compared to previous studies (Song et al. 2013; Citi et al. 2014; Ramirez and Paninski 2014), the numerics in our method lend themselves to an efficient implementation on both CPUs and GPUs. Moreover, the estimation procedure is highly amenable to parallelization, which makes it possible to scale up the number of units and putative connections dramatically.

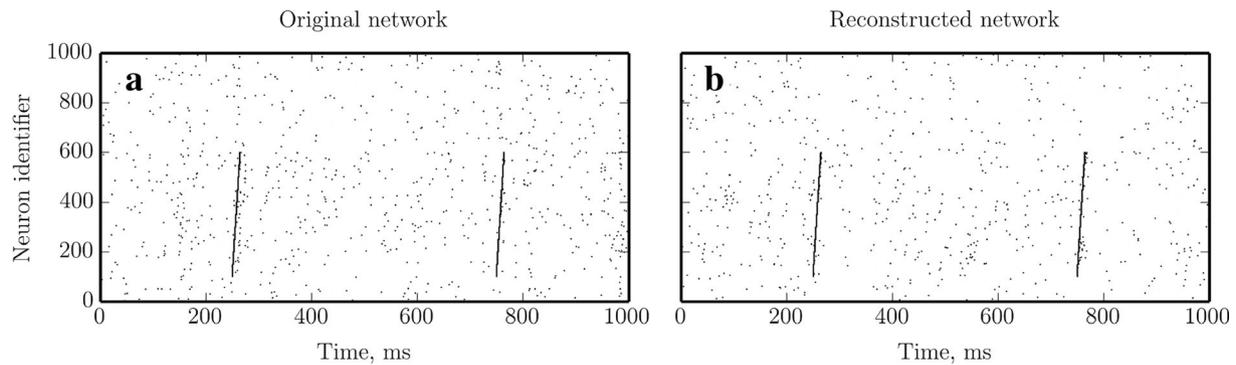

**Fig. 10** Synfire dynamics of the original and reconstructed networks. The first links in the synfire chains in both original and reconstructed network were stimulated (by injection of a strong input current) at 250 ms and 750 ms. The stimulation was performed using a large current pulse injection into all neurons of the first link. **a** Original network. **b** Reconstructed network

The proposed estimation procedure operates in continuous time on precise timestamps of the events (spikes), and does not require discretization, binning or smoothing of the data, which avoids the associated choice of bin or kernel size and induced artifacts (Ba et al. 2014). Additionally, unlike pairwise methods such as the coupled escape rate model (CERM) by Kobayashi and Kitano (2013), the reconstruction takes into account the complete ensemble spike history and thus is able to disambiguate complex indirect neural interactions. Other recently proposed connectivity reconstruction methods, not based on GLMs, exploit specific properties of leaky integrate-and-fire neurons (Van Bussel et al. 2011; Memmesheimer et al. 2014) or of linearly interacting point processes (Pernice and Rotter 2013). While this might be less clear for these methods, our procedure, since it is a MLE of a GLM, can be shown to have the optimality properties of becoming an asymptotically unbiased, consistent and efficient estimator of the ground truth connectivity in the limit of large sample sizes (Pawitan 2001) (provided that the suggested model is appropriate to describe neuronal dynamics). Moreover, it is amenable to efficient optimization via gradient ascent, since it is mathematically guaranteed to converge to the global maximum of the likelihood.

We present benchmarks against simulated random balanced networks of $N = 1000$ neurons with known ground truth connectivity, and show that our method achieves good performance for realistic model parameters and plausible amounts of data. Additionally, we performed a successful reconstruction of a structured network, where a synfire chain was embedded in a balanced network of excitatory and inhibitory neurons. The simulation of the reconstructed network with stimulation applied to the first link of the synfire chain, which was identified by carrying out cluster analysis of the recovered synaptic connectivity matrix, highlighted the generative properties of the GLM and showed virtually identical network dynamics to the original network. The application of cluster analysis to the reconstructed connectivity of the synfire chain is an example of how an inferred network model can be subdivided into interacting populations of neurons. Given such a partition of the network in functional subgroups, the activity dynamics can be analyzed using theory of population dynamics of GLM neurons (Deger et al. 2014).

Ideally our approach would also be validated against experimental data. Unfortunately, no datasets are currently available that contain long recordings of many individual spike trains and also the connectivity between the neurons. Indeed, generating such a dataset, although now technically possible, for example, using a high density microelectrode array setup (Ballini et al. 2013), would require extraordinary investment from an experimental laboratory. It is therefore more realistic to hope that the experimental validation of our technique can take place opportunistically on a dataset that is obtained for some other purpose.

In spite of the apparent simplicity of our model, the point process GLM framework that we used is very flexible and can be readily extended with additional features. The exponential kernels that we chose to describe the membrane filtering and nonlinear properties of the neurons can be replaced with more elaborate ones. For instance, previous works have represented neuronal interaction kernels by cosine "bumps" (Pillow et al. 2005), or composition of basis functions, such as Laguerre polynomials or B-splines (Song et al. 2013). However, in order to enable the reconstruction of networks of thousands of units, the key is to use functions that can both guarantee the concavity of the likelihood, as discussed in Paninski (2004), and at the same time make it possible to find analytical closed forms for the resulting expressions to enable efficient evaluation. These considerations, and the notion that the exponential PSP is a coarse first-order approximation to the dynamics of synaptic transmission, were the primary motivations for us to adopt the exponential kernels in this work. However, in Section 3.2 we have demonstrated that this simplification does not affect the reconstruction performance for the data generated by a more complex and realistic LIF model with $\alpha$-shaped PSCs. Besides, we argue that our model would be most useful to investigate network effects, as opposed to the effects explained by intricate features of the dynamics of individual synapses, for which purpose, conversely, smaller-scale, but more detailed models like the one by Song et al. (2013) might be more appropriate. Apart from that, it is possible to add supplementary terms to the membrane potentials of the neurons $U_i(t)$. One such option is to incorporate known external inputs directly into the model, such as those occurring in experimental paradigms widely used for studying predominantly stimulus-driven circuits like the retina (Pillow et al. 2008). Another option is to add unknown, common external inputs (Kulkarni and Paninski 2007; Vidne et al. 2012) in order to treat non-stationarity in the data.

A further possibility to improve the results of the estimation lies in enforcing Dale's law: neurons can be either inhibitory or excitatory, and they cannot form connections of both types at the same time (Eccles 1976). Unfortunately, the mathematical re-formulation of this law in the context of our model (the sign of all elements in each column of the synaptic weight matrix $J_{ij}$ should be identical) turns the original problem into a non-convex and non-separable one. Instead of trying to solve this much more difficult optimization problem, an approximate, greedy method can be implemented as outlined in Mishchenko et al. (2011). This involves first solving the original problem, then classifying the neurons as excitatory, inhibitory or unassigned based on the discovered synaptic weights, and, finally, imposing corresponding box constraints on the relevant elements of the $J_{ij}$ rows, which neither compromises the convexity, nor the separability properties. However, in our case, the major source of errors is the overlap between the unconnected and excitatory distributions, which generates non-Dale connections as a consequence. The benchmarks that we conducted show that very few of the errors are of the non-Dale category (see Tables 2, 3, 5), so any gain from imposing a Dale condition would be minimal and does not justify the additional complexity incurred. Therefore, effort should primarily be focused on tightening distributions.

In our GLM, we have used the exponential link function to map the membrane potential $U_i(t)$ to the instantaneous firing rate $\lambda_i(t)$. The exponential function is the canonical link function for the Poisson distribution, and it is commonly used in the single neuron modelling context, e.g. in the spike response model (Gerstner et al. 2014). Further reasons for us to choose an exponential function as the link function were as follows: 1) it has been previously shown (Jolivet et al. 2006; Mensi et al. 2012) that an exponential function is a good model for the nonlinear relationship between the conditional intensity of spike emission and the distance from the voltage threshold; 2) an exponential nonlinearity satisfies the sufficient condition established in Paninski (2004) for the likelihood of the model to be concave; 3) this choice makes it possible for us to obtain the closed form for the likelihood function as an exponential integral Ei (10), which is crucial here for reasons of computational efficiency. If the closed form cannot be obtained, then one either needs to discretize the likelihood integral, possibly using clever corrections to improve the accuracy (Citi et al. 2014), or, if the conditional intensity function can be shown to be piecewise smooth like in our case, attempt to get better precision and performance by applying quadrature methods to the smooth segments (Mena and Paninski 2014). Unfortunately, both approaches are still not fast enough for large GLMs such as ours. Other link functions such as logit and probit functions are also commonly used in the context of GLMs and have the property of being bounded (Song et al. 2013). Indeed, within our framework, it is possible to adopt a different link function instead of the canonical one. However, both logit and probit functions in particular are ruled out by the concavity condition (Paninski 2004), being saturating ("sigmoidal") nonlinearities. In practice, however, we did not experience any substantial issues due to the exponential function being positively unbounded. The box constraints that we imposed on the base rate and synaptic weights served only to repel the optimizer from the borders of the feasible region, where it might occasionally find itself due to an unfortunate combination of numerical artifacts. In none of the results presented in the paper did the recovered parameters turn out to be equal to the values of the box constraints.

Throughout this study, we have assumed that we have simultaneous access to all the spike trains of a neuronal population. For this case, and for a small number of neurons, it was shown that connectivity estimation via GLM can recover anatomical connectivity (Gerhard et al. 2013), as opposed to other methods, which mostly uncover "functional" or "effective" connections (Stevenson et al. 2008) that do not necessarily correspond to real synapses.

Here, we scaled the GLM approach up to large networks. However, in many experiments, such as in cortical multi-electrode array recordings (Truccolo et al. 2010), a complete recording of all neurons in a network is not feasible, but rather only parts of a neuronal network can be observed. With respect to the inference of connectivity from activity this is known as the problem of undersampling (Kim et al. 2011; Gerhard et al. 2011; Shimazaki et al. 2012; Lütcke et al. 2013): an unobserved neuron might excite several observed ones reliably and frequently. Even if these observed neurons are not synaptically connected to each other, connectivity inference methods that do not account for hidden units would infer connections among them to explain the correlations in their activity. Thus, we generally expect the reconstruction accuracy of our method to decrease in case of undersampling of the network, as the input from unobserved units will be "explained" by non-existing connections (see also Appendix C.2). Other experimental preparations, such as neuronal cultures on substrate-integrated multi-electrode arrays, are amenable to more complete recordings (Ballini et al. 2013), possibly enabling direct interpretation of the recovered connectivity.

We emphasize that our method is practical for networks of up to thousands of neurons, and yet we recognize that the machines featuring a large number of cores ($> 10^5$), such as the ones we used during the development phase of this project, are generally only to be found at major research institutions. These supercomputing facilities are becoming increasingly available to neuroscience researchers. For example, researchers based in Germany may take advantage of the twice-yearly calls for applications for computing time on the supercomputers at Jülich Supercomputing Centre[7], at no cost to the researcher if accepted. European researchers outside Germany can apply analogously for resources through PRACE,[8] and labs based in the US can apply for time at the NSF facilities.[9] Additionally, initiatives such as the Human Brain Project[10] and the Neuroscience Gateway[11] aim to make such resources more accessible to the neuroscience community. Even so, core-hour allocations often require a thorough justification and quantitative evidence of the scaling properties of the algorithm, both of which entail significant investment from the researcher in preparing the application.

Therefore, we also investigated the option of offloading the computations to the kinds of GPGPU accelerators that are currently available off-the-shelf. We implemented a naive version of a GPU port, in which the computation kernels originally written in C++ and parallelized using OpenMP to use multiple threads were rewritten using CUDA technology by Nvidia Corporation to use a GPU instead. In order to assess the performance of this port, we measured the time it takes to complete the reconstruction of the incoming synapses of one neuron of a network of a thousand of neurons, such as those presented in Section 3. Both applications were tested on an IBM System x iData-Plex dx360 M4 machine featuring two Intel Xeon X5650 processors (6 cores, 12 threads) and one NVIDIA Tesla M2070 (Fermi microarchitecture). The CPU version took 38 minutes to converge in 433 iterations, while the GPU port required 49 minutes and 427 iterations; the obtained log-likelihood values were identical up to an absolute difference of $\simeq 4 \times 10^{-10}$ and a relative difference of $\simeq 3 \times 10^{-15}$. This way, the speedup achieved by offloading the calculations to a single GPU as compared to a single CPU thread amounted to approximately a factor of 18. However, profiling revealed, that around 70 % of the runtime of the GPU port was not actually spent doing useful calculations, but rather transferring $v_{ij}$ vectors from the CPU to the GPU memory. Therefore, simply switching to a better GPU, such as the ones based on the Kepler microarchitecture, providing double of the data transfer bandwidth as compared to Fermi, will increase the speedup for a naive GPU port up to a factor of 28. Furthermore, we are currently investigating algorithmic improvements that completely remove the need for data transfers by storing $v_{ij}$ vectors directly in the GPU memory using specialized compression. Extrapolating on the performance from the proof-of-concept kernels we implemented, a future GPU realization may perform at least as fast as $\sim 55$ generic CPU threads, and require only a fraction of RAM as compared to the CPU-only realization by storing all of the working data in the onboard GPU memory. The complete GPU port of the method, along with its core algorithms and performance benchmarks, will be described in detail in a separate publication. The development of a substantially more efficient implementation will enable us to thoroughly investigate the limits of our approach in a way that is out of scope in the current study due to computational expense. One obvious area for investigation would be the degradation of performance in the case of undersampling as discussed above. Other areas worthy of further examination are the effects on misclassification error rates of correlated external inputs and non-stationarities in the recorded activity.

It is also important to mention that anatomically, cortical neurons receive on the order of $10^3$-$10^4$ incoming synapses (Braitenberg and Schüz 1991). In our demonstrations, we assumed that the network might be fully connected, or, in other words, each of the $N = 1000$ neurons can possibly receive up to $10^3$ incoming synapses from all other neurons, yielding $\mathcal{O}(N^2 = 10^6)$ parameters to constrain in total. However, given a substantially larger amount of recording channels, such as $N > 10^4$, if such data becomes available, this assumption is no longer reasonable. Instead, the data can be pre-processed to purge unlikely incoming connections, from $N$ down to the most likely $10^3$-$10^4$ putative synapses per neuron, thereby avoiding the quadratic explosion in the number of model parameters. We suggest that such pre-processing can be performed using computationally efficient pairwise linear methods, such as cross-correlation or cross-coherence, or various information theory metrics (Staniek and Lehnertz 2008). This way, while recovering the connectivity of even larger networks would still require a linear increase in computational resources (or wallclock time), the challenge to further scale the model up to a larger number of putative incoming synapses can be alleviated.

Finally, we would like to stress that even though network models that can be directly simulated as extracted from the data are interesting in themselves for further studies, the proposed method also has potential to provide insights into the network-wide plasticity of synaptic connectivity. Even though in our method we assume that the connectivity is fixed over the time of a recording, estimated synaptic weights can be tracked across several recordings performed in a time-lapse fashion. Such data could be relevant for models of synaptic plasticity over long time-scales (structural plasticity) (Escobar 2008; Deger et al. 2012), which currently have to rely on statistics of synapse numbers without temporal information, or time-lapse imaging of small numbers of individual synapses.

**Acknowledgments** We thank Stefano Cardanobile, Stefan Rotter, Wolfram Schenck and Christian Pozzorini for helpful discussions, and Ulrich Egert for his support of the initial phase of the project. We are grateful to Andrew V. Adinetz for his advice on analyzing the performance of our optimizer and on the GPU kernel implementation, and to Tom Tetzlaff for providing an example syn-fire chain simulation script. We acknowledge partial support by the German Federal Ministry of Education and Research (BMBF grants 01GQ0420 to BCCN Freiburg and 01GQ0830 to BFNT Freiburg), the Swiss National Science Foundation (grant agreement no. 200020_147200), and the Helmholtz Alliance through the Initiative and Networking Fund of the Helmholtz Association and the Helmholtz Portfolio theme "Supercomputing and Modeling for the Human Brain".

# Appendix

The spike data used in this paper and the code that implements our connectivity reconstruction method are publicly available for download at http://dx.doi.org/10.5281/zenodo.17662 and http://dx.doi.org/10.5281/zenodo.17663 respectively.

## A: Concavity of the point process log-likelihood

The derivative of the log-likelihood $\mathcal{L}_i$ (2) with respect to $J_{ij}$ is

$$\frac{\partial \mathcal{L}_i}{\partial J_{ij}} = \sum_{k=1}^{q_i} v_{ij}(t_{i,k}) - \int_{T_0}^{T_1} \lambda_i(t) v_{ij}(t) dt \, ,$$

cf. (12), with $v_{ij}(t) = (h_i * s_j)(t)$ and $v_{i0} = 1$ (11). The second derivative is then simply

$$\frac{\partial}{\partial J_{ik}} \frac{\partial}{\partial J_{ij}} \mathcal{L}_i = - \int_{T_0}^{T_1} \lambda_i(t) v_{ij}(t) v_{ik}(t) dt \, . \tag{18}$$

Using these expressions, in the following we give a proof, specific to our model, that $\mathcal{L}_i$ is concave. A condition for the concavity of the log-likelihood of more general point process GLMs is given in Paninski (2004).

A twice differentiable function of several variables is concave if and only if its Hessian matrix $\mathcal{H}$ is negative semi-definite. In terms of the parameter vector $\theta_i = (J_{i0}, \ldots, J_{iN})$, the Hessian matrix of $\mathcal{L}_i$ has the elements (18). This matrix is negative semi-definite if $x^T \mathcal{H} x \leq 0$ for all real vectors $x$. We evaluate this expression as

$$x^T \mathcal{H} x = \sum_j x_j \sum_k \mathcal{H}_{jk} x_k = \sum_{j,k} x_j x_k \frac{\partial}{\partial J_{ik}} \frac{\partial}{\partial J_{ij}} \mathcal{L}_i$$

$$= - \int_{T_0}^{T_1} \lambda_i(t) \sum_k x_k v_{ik}(t) \sum_j x_j v_{ij}(t) dt$$

$$= - \int_{T_0}^{T_1} \lambda_i(t) V_i^2(t) dt \leq 0 \, , \tag{19}$$

where $V_i(t)$ is defined analogously to $U_i(t)$ (3) as $\sum_{j=0}^{N} x_j v_{ij}(t)$, and $\lambda_i(t) \geq 0$. Hence $\mathcal{H}$ is negative semi-definite, and thus $\mathcal{L}_i$ is concave in the parameters $\theta_i$.

## B: Closed form for the log-likelihood integral

To evaluate Eq. (9), we need to compute the term

$$\int_{t_k}^{t_{k+1}} \exp \left\{ (U_i(t_k) - J_{i0}) e^{-\frac{t-t_k}{\tau_i}} \right\} dt \, .$$

Let us introduce the shorthand $g(t) = (U_i(t_k) - J_{i0}) e^{-\frac{t-t_k}{\tau_i}}$. We need to show that the exponential integral $\mathrm{Ei}(x) =$ $-\int_{-x}^{\infty} \frac{e^{-t}}{t} dt$ is a primitive of $\exp\{g(t)\}$ for $t_k \leq t < t_{k+1}$. We differentiate

$$\frac{d}{dt} \mathrm{Ei}(g(t)) = -\frac{d}{dt} \int_{-g(t)}^{\infty} \frac{e^{-u}}{u} du = \frac{e^{g(t)}}{-g(t)} \left( -\frac{d}{dt} g(t) \right)$$

$$= -\frac{1}{\tau_i} e^{g(t)}.$$

Thus, we can evaluate the integral as $\int_{t_k}^{t_{k+1}} \exp\{g(t)\} dt = -\tau_i \, \mathrm{Ei}(g(t))|_{t_k}^{t_{k+1}}$, and so follows (10).

## C: Spot checks for several degrees of undersampling and sparsity

### C.1: Chance level of the misclassification error rate

Connections in our networks are formed with a connection probability $p$. A fraction $f_e$ of neurons is excitatory, the remainder $f_i = 1 - f_e$ is inhibitory. To assess the quality of our connectivity reconstruction, here we compute the misclassification error rate (MER) of a random connection classifier that maintains $p$, $f_e$ and $f_i$. We call this the chance level $\mathrm{MER}_0$.

Misclassification errors can occur for three types of synaptic connections: excitatory, inhibitory and null connections. Let us take the example of the excitatory type. We expect $p(f_e N)(N-1)$ excitatory connections, each of which is misclassified (false negative) with probability $(1 - pf_e)$, because with probability $pf_e$ it would be classified correctly as excitatory. Analogously the expected number of misclassifications of each type is given by

- excitatory: $n_e = p(f_e N)(N-1)(1 - pf_e)$;
- inhibitory: $n_i = p(f_i N)(N-1)(1 - pf_i)$;
- null: $n_n = (1-p)N(N-1)p$ .

The total rate of errors is then the expected number of errors, summed over types, divided by the total number of possible connections. This yields the following expression:

$$\mathrm{MER}_0 = \frac{n_e + n_i + n_n}{N(N-1)} = p\left(2 - p(1 + f_e^2 + f_i^2)\right) , \quad (20)$$

which is independent of $N$, but depends on the connection probability $p$ and the ratio of excitatory to inhibitory neurons.

### C.2: Effects of undersampling

To assess the degree to which undersampling deteriorates the quality of the network reconstructions, we performed several experiments with different datasets, each being a subsample of the original one presented in Section 3.1. In each experiment we randomly selected a fraction of neurons (maintaining the ratio of excitatory and inhibitory neurons)

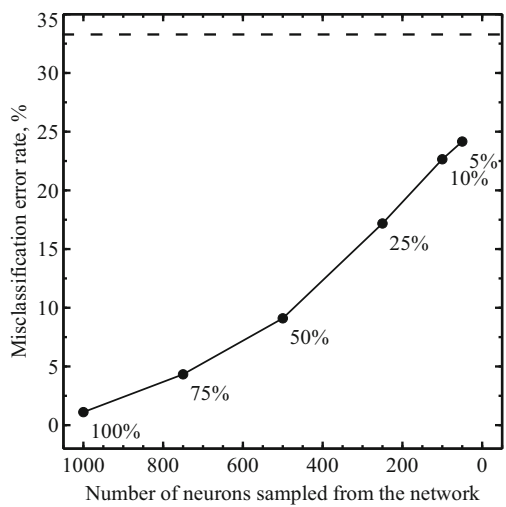

**Fig. 11** Misclassification error rate as a function of the number of sampled neurons. The reconstructions on the partial datasets were performed with the parameters and data presented in Section 3.1 (*black dots*). The full network consists of $N = 1000$ neurons, partial datasets consist of the spike trains of a smaller number of randomly selected neurons (maintaining the ratio of excitatory and inhibitory cells). Annotations give the percentage of neurons sampled from the full network. The chance level MER (20) was calculated for $p = 0.2$, $f_e = 0.8$ and $f_i = 0.2$ ($\mathrm{MER}_0 = 33.28$ %; *dashed line*). The synaptic weights were classified using $k$-means as described in Section 3.3

that are fed into the optimizer. The results are shown in Fig. 11. In contrast to Fig. 3, here the connections were classified using $k$-means as described in Section 3.3, which is more robust in the undersampled cases. Therefore, for the case of $N = 1000$ neurons the MER is slightly higher than when classified using GMM, as reported in Table 2. As expected, the MER of the partial network increased as we decreased the number of neurons that were visible to the GLM (undersampling). This was largely due to the broadening of the distribution of the synaptic weights for null connections (data not shown, cf. Figure 3). Yet, in all cases, synapse classification based on the reconstruction method was substantially better than random classification of the synapses, see Appendix C.1 for the derivation of the chance level MER.

**Table 6** Breakdown of classification errors for the GLM random network with sparsity $p = 0.1$

| Connection type | Errors | FP | FN | ND |
|---|---|---|---|---|
| Excitatory | 2 554 | 67 % | 33 % | 19 % |
| Inhibitory | 0 | | | |
| Unconnected | 2 554 | 33 % | 67 % | — |
| Total errors | 0.26 % | | | |

The meaning of the abbreviations is the same as in Table 2

## C.3: Effects of varying connection sparsity

In this experiment we performed the reconstruction on a dataset simulated as described in Section 3.1, but with connection probability of $p = 0.1$ instead of $p = 0.2$. The results are presented in Table 6. Note that whereas the quality of the reconstruction is substantially better than for $p = 0.2$ (shown in Table 2), the chance level of the misclassification error rate for this network with $p = 0.1$ is $MER_0 = 18.32$ %, rather than $MER_0 = 33.28$ % for the network with $p = 0.2$. Still, also in relative terms to $MER_0$, the reconstruction is more accurate in this case of increased connection sparsity.